\documentclass[12pt]{article}
\usepackage[dvips]{graphicx}
\usepackage{epsfig,cite}
\usepackage {amsmath}
\usepackage{amssymb}
\usepackage{slashed}
\usepackage{cancel}
\usepackage{lscape}

\def\beq{\begin{equation}}
\def\eeq{\end{equation}}

\begin{document}

\begin{titlepage}
\pagestyle{empty}
\rightline{UMN--TH--3211/13, FTPI--MINN--13/21}
\vskip +0.4in

\begin{center}
{\large {\bf Affleck-Dine Baryogenesis and Inflation  in Supergravity with Strongly Stabilized Moduli}}
\end{center}

\begin{center}
\vskip +0.25in
{\bf Marcos A.G. Garcia$^1$ and Keith A. Olive$^2$}
\vskip 0.2in
{\small {\it
$^1${School of Physics and Astronomy,\\
University of Minnesota, Minneapolis, MN 55455,\,USA} \\
$^2${William I. Fine Theoretical Physics Institute, School of Physics and Astronomy,\\
University of Minnesota, Minneapolis, MN 55455,\,USA}
}}


\vspace{0.8cm}
{\bf Abstract}
\end{center}
{\small Constructing models of inflation and/or baryogenesis in the context of 
$\mathcal{N}=1$ supergravity is known to be difficult as the finite energy density during inflation
typically generates large (order the Hubble scale) mass terms. This is the well-known
$\eta$ problem in inflation. The same effect gives masses along 
low energy flat directions of the scalar potential thus potentially preventing
Affleck-Dine baryogenesis to occur. It has been shown that adding a chiral multiplet
$S$ coupled to the inflaton (with a shift symmetry) can serve to stabilize the inflationary potential and allows one to derive simple inflationary potentials without an $\eta$ problem.  Here, we show that
by coupling the {\em same} stabilizing field $S$ to the flat direction, may naturally 
lead to a negative mass-squared contribution to the flat direction
thus generating the necessarily large vacuum expectation value needed to realize
Affleck-Dine baryogenesis. We trace the evolution of the inflaton, stabilizer, and flat direction field,
as well as a Polonyi-like modulus responsible for soft supersymmetry breaking.
}


\vfill
\leftline{June 2013}
\end{titlepage}


\section{Introduction}

The evidence for a baryon asymmetry in the universe strongly suggests the need for new physics beyond the Standard Model. There are of course a number of viable mechanisms to account for the observed ratio of baryons to photons \cite{wmap}, 
\begin{equation}
\frac{n_B}{n_{\gamma}}= 6.05 \pm 0.08 \times 10^{-10}
\end{equation}
including GUT baryogenesis, supersymmetric electroweak baryogenesis and leptogenesis. However, the out-of-equilibrium decay of coherent scalar fields associated with supersymmetric flat directions, known as the Affleck-Dine (AD) mechanism \cite{AD}, stands as one of the most attractive scenarios. Within the MSSM, many of these ($D$- and $F$-) flat directions carry non-zero baryon and/or lepton numbers. 
If these flat directions, which we will denote collectively as $\Phi$, are excited during inflation,
their subsequent evolution may lead to a net baryon asymmetry.  So long as 
soft supersymmetry breaking masses which lift the flatness are small compared with the Hubble parameter, these excitations are unavoidable, moderated only by possible non-renormalizable terms in the superpotential.  After inflation, once the Hubble parameter becomes comparable to the effective mass of $\Phi$, oscillations of $\Phi$ start about the true minimum. $B-L$ and $CP$ violation  then lead to a large non-zero asymmetry during these oscillations. 

Implementation of AD baryogenesis in the context of supergravity can be problematic
because of a problem reminiscent of the $\eta$ problem in inflation \cite{eta,eds}. 
In $\mathcal{N}=1$ supergravity with minimal kinetic terms, a large vacuum energy density, $V\sim H_I^2 M_P^2$  where $H_I$ is the Hubble parameter during inflation, and $M_P$ is the Planck mass, typically
leads to an induced mass for all scalar fields of order $H_I$ \cite{glv}. If the inflaton also receives an
order $H_I$ contribution to its mass, the slow roll parameter $\eta$ will be $\mathcal{O}(1)$ and
inflation will not occur. A similar problem may occur for AD baryogenesis if the flat directions
receive masses of order $H_I$ during inflation \cite{DRT}. In this case, no excitation of the 
flat direction occurs during inflation and no baryon asymmetry is produced.

Of course inflationary models have been constructed in the context of $\mathcal{N}=1$ supergravity
\cite{nost}, by making use of an accidental \cite{lw} cancellation in the superpotential
or by modifying the K\"ahler potential. In particular, no-scale supergravity \cite{no-scale}
has been particularly helpful in the construction of inflationary models \cite{noscinfl,eds}.
It is also possible to construct models based on a more general Heisenberg symmetry
\cite{BG}.

AD baryogenesis will occur if the contribution to the soft mass squared
of the flat direction during inflation is negative \cite{DRT}.  While the correction is easily
seen to be positive in minimal supergravity, it is possible that with a suitable modification of the
K\"ahler potential, a negative contribution is possible \cite{ds}.  
Another possibility to generate negative effective masses along the flat direction, is to consider non-minimal K\"ahler terms coupling the inflaton and the flat direction \cite{DRT}. However, the form of these terms and the constraints that must be imposed on them depend on the specific model of inflation. 
A number of superstring inspired examples can be found in \cite{Higaki}.
Most studies of AD baryogenesis simply
assume a negative mass-squared contribution. (For reviews on AD baryogenesis see: \cite{reviews}.)  
In this case, too, no-scale supergravity may
rescue the situation as the tree level supergravity contribution to the mass squared is identically 
zero (this is true for any K\"ahler potential respecting the Heisenberg symmetry)
and 1-loop corrections are typically negative  \cite{gmo,cgmo}.

There is another class of inflation models which avoids the difficulty with the $\eta$ problem by utilizing a shift symmetry for the inflaton in the K\"ahler potential \cite{KYY}. This idea was used in several subsequent papers \cite{other,DP,KLR,KLOR}.The field content of the theory consists of the inflaton $I$ and a stabilizer field $S$ that is forced to vanish during inflation. The shift symmetry is defined so that the K\"ahler potential is chosen to be independent of the combination $I+\bar{I}$, which plays the role of the inflaton. Therefore, the dangerous term stemming from $e^K$ is absent, making the potential flat. By requiring the superpotential to vanish during inflation, the potential can be taken to be any specific function of the inflaton.

Here, we will consider an alternative mechanism for the generation of the negative masses squared along flat directions, which relies on the coupling through non-minimal K\"ahler couplings between the flat direction and a stabilizer field. Because there is no direct coupling between the inflaton and the flat direction, our results are by an large independent of the specific model of inflation.
Put another way, our results are applicable to any single field inflationary model,  formulated 
through a shift symmetry and stabilizer field \cite{KYY}.
In particular, they will allow the direct realization of the Affleck-Dine scenario within chaotic inflation. 
We also include in our analysis the effects of a strongly stabilized Polonyi-like modulus which
is responsible for generating soft supersymmetry breaking.
After some preliminary set-up of the field content in section 2, we introduce the coupling between
a generic flat direction and the stabilizing field $S$ and derive the resultant scalar potential in section 3.
Our numerical results are given in section 4 and concluding remarks in section 5.

\section{Field Content}

We will work in the frame work of the minimal supersymmetric standard model.
 The scalar potential for uncharged chiral superfields is \cite{Fetal}
\begin{equation}\label{sugra_potential}
V=e^{K}(K^{i\bar{j}}D_{i}W\bar{D}_{\bar{j}}\bar{W}-3|W|^2),
\end{equation}
where $K$ is the K\"ahler potential, $W$ is the superpotential, and
\begin{equation}
D_{i}W=K_{i}W+\partial_{i}W.
\end{equation}
The reduced Planck mass, $M_P$, has been set to be unity.
We assume that the low energy theory is derivable from an $\mathcal{N}=1$ supergravity theory in
the decoupling limit. 

A hidden sector with a Polonyi type superpotential \cite{pol} will be considered as the source of soft supersymmetry breaking. Specifically, we consider a non-minimal Polonyi model based on the so-called O'KKLT mechanism \cite{dine}. The Polonyi sector includes a strong stabilizing term added to an otherwise minimal K\"ahler potential,
\begin{equation}
K_{\text{\cancel{susy}}} = Z\bar{Z}-\frac{(Z\bar{Z})^2}{\Lambda^2},
\label{okklt}
\end{equation}
where it is assumed that $\Lambda\ll 1$. The superpotential is the ordinary Polonyi superpotential
\begin{equation}\label{w_z}
W_{\text{\cancel{susy}}} = \mu^2(Z+\nu),
\end{equation}
where the parameter $\nu$ must be tuned to yield a vanishing cosmological constant when $Z$ sits at the minimum of the potential.  The Polonyi field $Z$ will be parametrized as
\begin{equation}\label{zeta}
Z=\frac{1}{\sqrt{2}}(z+i\chi).
\end{equation}
This choice of K\"ahler potential and superpotential provides the strong stabilization for $Z$
which has a real minimum ($\langle \chi \rangle =0$), and 
\begin{equation}\label{z_min}
\langle z \rangle \simeq\frac{\Lambda^2}{\sqrt{6}}\ , \quad \nu\simeq \frac{1}{\sqrt{3}}.
\end{equation}
The supersymmetry breaking mass scale given by the gravitino mass is 
\beq 
m_{3/2}=e^{K_{\text{\cancel{susy}}}/2}\langle W_{\text{\cancel{susy}}} \rangle \sim\mu^2/\sqrt{3} \, ,
\eeq
whereas the mass squared of both $z$ and $\chi$ are $m_{z,\chi}^2\sim m_{3/2}^2/\Lambda^2\gg m_{3/2}^2$. Therefore, $Z$ is stabilized, and for a sufficiently small $\Lambda$ the cosmological moduli and gravitino problems \cite{prob1,glv,prob2} can be resolved \cite{sol,lmo,DLMMO,bdd}. In this case, dilution from moduli decay is minimized \cite{cgmo,bdd,kane}

In principle, we could also include in the hidden sector a volume modulus $\rho$, necessary in string theory to ensure that the universe is 4d. The dynamics of the modulus $\rho$ will not be analyzed here. However it was shown that if the volume modulus is stabilized by using the KL superpotential \cite{KL},  this field is relatively unperturbed when one includes the inflationary sector considered (see \cite{KLOR}).
For details on strongly stabilized theories involving both a Polonyi sector based on Eq. (\ref{okklt}) and
a volume modulus stabilized with a KL superpotential see \cite{DLMMO}. 

The inflationary sector of the model consists of an inflaton $I$ and a stabilizer field, $S$, which are the necessary ingredients for the theory of inflation of  \cite{KYY,DP,KLR,KLOR}. We will write them in the basis
\begin{equation}\label{inffields}
S=\frac{1}{\sqrt{2}}(s+i\alpha)\ , \quad I=\frac{1}{\sqrt{2}}(\eta+i\beta).
\end{equation}
The real part of $I$, $\eta$, will play the role of the inflaton, while $S$ and $\beta$ will be forced to vanish during inflation. Constraining the K\"ahler potential to be independent of $\eta$ ensures the flatness of the potential. We will specialize to the following form of the K\"ahler potential,
\begin{equation}
K_{\rm inf} = -\frac{1}{2}(I-\bar{I})^2 +S\bar{S} - \xi (S\bar{S})^2.
\end{equation}
For this choice of $K$, the fields $S$ and $I$ are canonically normalized along the inflaton path $S=0$, $\beta=0$.  The superpotential is taken as
\begin{equation}\label{w_inf}
W_{\rm inf}=Sf(I),
\end{equation}
where $f$ is a real holomorphic function. The scalar potential (\ref{sugra_potential}) is an even function of $S$, and must be invariant with respect to $I\longrightarrow \bar{I}$, making it even as a function of $\beta$ . Therefore, the potential has an extremum along the inflaton path, and it takes the remarkably simple form
\begin{equation}
V (\eta) = f^2(\eta/\sqrt{2}).
\end{equation}
For positive $\xi$ this extremum is a minimum. The condition for the stability of this minimum under quantum fluctuations is $m_{\perp}^2\geq H^2$, where $m_{\perp}$ are the transverse masses. A straightforward calculation reveals it is satisfied for $\xi\gtrsim 1/12$; cf. \cite{KLR,KLOR}. In what follows we will specialize to the choice $f(I)=m_{\eta}I$, for which the inflationary potential takes the form
\begin{equation}
V(\eta)=\frac{1}{2}m_{\eta}^2\eta^2,
\end{equation}
i.e. that of quadratic chaotic inflation \cite{Linde:1983gd}. We stress that we use this only
as the simplest example for inflation and can easily substitute other forms of $f(\eta)$ 
which are perhaps more in accord with recent Planck results \cite{planck}. For other examples
of chaotic models see \cite{newY}.

The MSSM contains many $F$- and $D$-flat directions, along which the renormalizable scalar potential vanishes identically, in the supersymmetric limit. Realistically, these `flat' directions are only approximately flat, since they are lifted by soft supersymmetry-breaking terms, and by non-renormalizable terms in the superpotential \cite{catalog}. These flat directions allow gauge invariant combinations of squark and/or slepton fields to develop non-zero VEVs. For the MSSM superpotential,
\begin{equation}
W_{\rm MSSM}= y^{(u)}_{ij}Q^{a}_{i\alpha}\bar{u}_{ja}H^{\alpha}_{u} + y^{(d)}_{ij}Q^{a}_{i\alpha}\bar{d}_{ja}H^{\alpha}_{d} + y^{(e)}_{ij}L_{i\alpha}\bar{e}_{j}H^{\alpha}_{d} + \mu^{(H)} H_{u\alpha}H_{d}^{\alpha}
\end{equation}
one of these (approximately) flat directions, for which $B-L\neq0$, can be parametrized by the complex scalar field $\Phi$ as follows,
\begin{equation}
L_{1}=\left(
\begin{matrix}
\Phi\\
0
\end{matrix}
\right)\ , \quad H_{u}=\left(
\begin{matrix}
0\\
\Phi
\end{matrix}
\right).
\end{equation}
A complete catalog of MSSM flat directions can be found in \cite{catalog}. In addition to the soft supersymmetry breaking terms to be discussed in more detail below, this particular flat direction can be lifted by a non-renormalizable term of the form
\begin{equation} \label{AD_W}
W_{\rm AD} = \frac{\lambda}{M}(L_{1\alpha}H_{u}^{\alpha})^2 = \frac{\lambda}{M}\Phi^4
\end{equation}
where $M$ is some large mass scale, such as the GUT or Planck scale. The non-renormalizable contribution might arise directly at the string scale or be generated by integrating out a heavy scalar singlet $N$ with coupling $y^{(N)}L_{\alpha}H_{u}^{\alpha}N$. For simplicity, our discussion of the AD sector of the theory will be chosen to consist of the single flat direction with superpotential (\ref{AD_W}). For more complex scenarios involving multiple flat directions see \cite{multiple}. We will refer to $\Phi$ as the AD field and write it in the following basis,
\begin{equation}\label{adparam}
\Phi=\frac{1}{\sqrt{2}}(\phi+i\gamma). 
\end{equation}
 The renormalizable theory has an approximately conserved current $j_{\mu}=i(\Phi^*\partial_{\mu}\Phi-\Phi\partial_{\mu}\Phi^*)$ and $CP$ invariance $\Phi\longleftrightarrow \Phi^*$. We will refer to $n_{B}=j_0=(\gamma\dot{\phi}-\phi\dot{\gamma})$ as the baryon number density \cite{AD,lin}. 
 Strictly speaking, with the flat direction described above, we only produce a lepton asymmetry
 \cite{cdo} which is subsequently converted to a baryon asymmetry through sphaleron interactions \cite{fy}.

With this field content and our choices of the K\"ahler potential and superpotentials, 
we can derive the resulting scalar potential from Eq. (\ref{sugra_potential}).

\section{AD Baryogenesis and Inflation}

In most supergravity models, the finite energy density during inflation breaks supersymmetry and induces an effective mass for $\Phi$ of the order of the Hubble parameter, $|m^2_{\Phi}|\sim H^2\gg m_{3/2}^2$. For minimal K\"ahler terms, this squared mass is positive, and the AD field is driven to zero, preventing the generation of a baryon asymmetry. It is often argued that non-minimal K\"ahler terms coupling the inflaton and the AD field are necessary to obtain a negative squared mass, which would allow $\Phi$ to obtain a non zero VEV during inflation. In our analysis, the presence of the stabilizer field $S$ in the inflationary sector allows for an alternative method for generating these effective masses, without significantly perturbing the dynamics of the inflaton. For a related analysis see \cite{kp}. In addition to the K\"ahler terms described above, the K\"ahler potential for the AD and S fields will be chosen as,
\begin{equation}
K_{\rm AD}=\Phi\bar{\Phi}+\zeta (S\bar{S})(\Phi\bar{\Phi}).
\end{equation}
Other matter fields will be assumed to have minimal K\"ahler terms.

Having set up the ingredients of the model, let us consider the evolution of the AD field. During inflation the Polonyi field $Z$ will be displaced from its true minimum given by Eq. (\ref{z_min}) to smaller values. 
Since the dominant contribution to the potential comes from the inflationary sector, the AD field can be ignored for the purpose of calculating the VEV of $Z$.  Let us consider the superpotential (\ref{w_inf}), which leads to 
\begin{equation}\label{zmininf}
\begin{aligned}
V&\simeq e^{K_{\text{\cancel{susy}}}(Z\bar{Z})}\left[|f(I)|^{2}+K^{Z\bar{Z}}D_{Z}W_{\text{\cancel{susy}}} \bar{D}_{\bar{Z}}\bar{W}_{\text{\cancel{susy}}} -3|W_{\text{\cancel{susy}}}|^2 \right]\\
&=e^{Z^2-Z^4/\Lambda^2}\left[f^2(\eta/\sqrt{2})+\mu^4\left( \frac{\left(1+Z(Z+\nu)(1-2Z^2/\Lambda^2)\right)^2}{(1-4Z^2/\Lambda^2)} -3(Z+\nu)^2 \right)\right]\\
&\simeq f^2+\mu^4(1-3\nu^2)-4\mu^4\nu Z + \left[f^2-2\mu^4(\nu^2-2/\Lambda^2) \right]Z^2+\mathcal{O}(Z^3).
\end{aligned}
\end{equation}
Since during inflation, $f^2\simeq 3H^2\gg \mu^4$, the expectation value for $Z$ following from (\ref{zmininf}) is
\beq
\langle Z\rangle _{\rm inf} \simeq \frac{2}{3}\left(\frac{\mu^2}{H}\right)^2\nu\simeq \frac{2}{3\nu}\left(\frac{m_{3/2}}{H}\right)^2 \ll 1
\label{zmininf2}
\eeq
Because the vev of $Z$ is small during inflation (i.e. much smaller than its final value), the gravitino mass has approximately the same value $\mu^2 \nu$ during inflation. 

For a generic non-renormalizable superpotential $W_{\rm AD}=\lambda\Phi^{n}/M^{n-3}$, 
 along the inflatio\-nary direction, $S=\beta=0$, the scalar potential takes the form
\begin{equation}\label{AD_potential}
\begin{aligned}
e^{-|\Phi|^2}V(\eta,\Phi) =&\ f^2(\eta/\sqrt{2})(1+\zeta|\Phi|^2)^{-1} + m_{3/2}^2|\Phi|^2 + Am_{3/2}\left(\frac{\lambda}{M^{n-3}}\Phi^n+h.c.\right) \\ 
& +n^2\frac{|\lambda|^2}{M^{2(n-3)}}|\Phi|^{2(n-1)} + m_{3/2}|\Phi|^2\left(\frac{\lambda}{M^{n-3}}\Phi^n + h.c.\right)\\
& + \left(2n-3+C\right)\frac{|\lambda|^2}{M^{2(n-3)}}|\Phi|^{2n}  + \frac{|\lambda|^2}{M^{2(n-3)}}|\Phi|^{2(n+1)}
\end{aligned}
\end{equation}
where
\begin{align}
A&=n-3+Z(Z+\nu)^{-1}\left(1+Z(Z+\nu)(1-2Z^2/\Lambda^2)\right)(1-2Z^2/\Lambda^2)(1-4Z^2/\Lambda^2)^{-1},\\
C&= Z^2(1-2Z^2/\Lambda^2)^2(1-4Z^2/\Lambda^2)^{-1}.
\end{align}
The `$A$-term' includes all contributions proportional to $W_{\rm AD}$. It is worth noting that there is no contribution from the inflationary sector to $A$; as a result, there is no $A$-term associated with the mass scale $H$. This can be tracked to the vanishing of the imaginary part of $I$. The `initial' phase of the AD field will then be essentially random, due to the de Sitter fluctuation of $\Phi$. At the minimum (\ref{z_min}), the condition of vanishing cosmological constant implies $3C=(A-n+3)^2$.
Because $\langle Z \rangle \ll 1$ (in Planck units), $A \approx n-3$ and $C \approx \langle Z \rangle^2$.

We will specialize to the case $n=4$, and $f^2(\eta/\sqrt{2})=m_{\eta}^2\eta^2/2M_P^2$. Restoring the Planck scale, $M_P$, for $|\Phi|^2<M_P$, Eq. (\ref{AD_potential}) reduces to
\begin{equation}\label{app_pot}
\begin{aligned}
V=&\ \frac{1}{2}m_{\eta}^2\eta^2+\left[m_{3/2}^2+(1-\zeta)\frac{m_{\eta}^2\eta^2}{2M_P^2}\right]|\Phi|^2\\
& + Am_{3/2}\left(\frac{\lambda}{M}\Phi^4+h.c.\right)+16\frac{|\lambda|^2}{M^2}|\Phi|^6+ \mathcal{O}(M_P^{-2}),
\end{aligned}
\end{equation}
with $A\simeq 1$. 
The inflaton dominates the energy density during inflation, $m_{\eta}^2\eta^2/2M_P^2\sim 3H^2\gg m_{3/2}^2$. Therefore, for $\zeta>1$, the effective mass of the AD field is negative, and therefore the field is displaced from the origin. The instantaneous minimum is located at
\begin{equation}\label{phi_min}
|\Phi_0|^2\simeq \left(\frac{\zeta-1}{96}\right)^{1/2}\frac{m_{\eta}\eta M}{|\lambda| M_P} \simeq (\zeta-1)^{1/2}\frac{HM}{4|\lambda|}.
\end{equation}

The addition of the AD and Polonyi fields to the inflationary sector perturbs the dynamics of the inflaton and the stabilizer field. This perturbation to the inflationary trajectory can be found expanding the complete scalar potential to quadratic order and solving for the displacement that minimizes it. 
A similar analysis was performed for the inflationary sector and a volume modulus \cite{KLOR}. 
Details for the present field content can be found in the Appendix.

At the end of inflation, when $\eta=\sqrt{8/3} M_P$, the inflaton starts oscillating about $\eta=0$. These coherent oscillations dominate the energy density of the universe, redshifting as pressureless matter, $H=\frac{2}{3}t^{-1}$. Denoting by $R_{\eta}$ the scale factor at the start of oscillations, we can write
\begin{align} 
\rho_{\eta} & = \frac{1}{2}  m_{\eta}^2\eta^2 = \frac{4}{3} m_{\eta}^2M_P^2\left(\frac{R_{\eta}}{R}\right)^3 \, , \\
H & =  \frac{2}{3}m_\eta \left(  \frac{R_{\eta}}{R}  \right)^{3/2} \label{h_eta}
\end{align}
During inflation, the vev of  $\Phi$ will be given by Eq. (\ref{phi_min}). 
Because its mass is of order $H$, the flat direction tracks its instantaneous minimum (\ref{phi_min})
as does the Polonyi field. Both are examples of the adiabatic relaxation discussed in \cite{adrel}. 
Since the field $S$ is stabilized, the AD field evolves according to the equation
\begin{equation}\label{phi_eom}
\ddot{\Phi}+\frac{2}{t}\dot{\Phi}+V'(\Phi)=0 \, ,
\end{equation}
however, the instantaneous minimum (\ref{phi_min}) decreases with time as $H^{1/2}\sim t^{-1/2}$.
As the amplitude of inflaton oscillations begins to damp due to Hubble expansion, one can show that the subsequent evolution of the AD field will track its minimum closely. 
Let us parametrize the AD field in terms of the instantaneous minimum and a dimensionless field $\sigma$, $\Phi(t)=\sigma(t) \Phi_0(t)$. The potential can be taken as (\ref{app_pot}), noting that during this era $H\gg m_{3/2}$. Equation  (\ref{phi_eom}) is then homogeneous in $t$. The substitution $t=e^{\tau}$ results in the following equation for $\sigma(\tau)$,
\begin{equation}
\sigma''-\left[\frac{1}{4}+\frac{4}{3}(\zeta-1)\right]\sigma+\frac{4}{3}(\zeta-1)\sigma^5=0,
\end{equation}
where prime denote the derivate with respect to $\tau$. 

This equation describes undamped oscillations about the fixed point $\sigma=[1+\frac{3}{16(\zeta-1)}]^{1/4}\sim 1$. During inflation, $H$ is approximately constant and the AD field sits at the instantaneous minimum, and thus $\sigma$ starts at the fixed point. $\Phi$ should then just track $\Phi_0$. However, during the transition from inflation to matter-like oscillations of $\eta$, the time dependence of the Hubble parameter changes. $\Phi$ may then overshoot the instantaneous minimum, in which case $\Phi$ oscillates about $\Phi_0$ with an envelope decreasing as $t^{-1/2}$, effectively tracking  the minimum. 

The AD field will start oscillations about the true minimum $\Phi=0$ when the Hubble parameter becomes comparable to the soft mass, $H=\frac{2}{3} m_{3/2}$. We must now follow the evolution of the two-field oscillating system \cite{eeno,gmo,cgmo}. The details of this stage depend on the moment at which reheating occurs. For definiteness, let us consider the mass (in the absence of the Hubble correction) of the AD field, $m_{\Phi}=m_{3/2}\sim 1\,{\rm TeV}\sim 10^{-15}M_P$. 

Reheating after inflation is determined by the couplings of the inflaton to matter
and surprisingly in this model, the decay probability of the inflaton is extremely small \cite{KLOR} re\-miniscent of the
situation in no-scale supergravity models \cite{ekoty}. In \cite{KLOR}, it was assumed that the gauge kinetic function
depends linearly on the inflaton, with coupling $d_\eta$. This allows for a gravitational-strength decay of the inflaton to two gauge bosons, for which the decay rate is expected to be
\beq 
\Gamma_{\eta}\simeq \frac{3d_{\eta}^2}{64\pi}\left(\frac{N_G}{12}\right)\frac{m_{\eta}^3}{M_P^2} \sim10^{-2} d_\eta^2 \frac{m_{\eta}^3}{M_P^2} \equiv {\tilde{d}_{\eta}}^2 \frac{m_{\eta}^3}{M_P^2}, 
\eeq
where $N_G$ is the number of final states, $N_G=12$ for the Standard Model.
Thus, in the ins\-tantaneous approximation, the inflaton decays when $\Gamma_{\eta}=\frac{3}{2} H$, or $R_{d\eta}/R_{\eta}= (M_P/\tilde{d}_{\eta}m_{\eta})^{4/3}$. After its decay, the universe is dominated by the relativistic decay products, with energy density and Hubble parameter
\begin{align}
\rho_r &= \frac{4}{3}\tilde{d}_{\eta}^{-4/3} m_{\eta}^{2/3}M_P^{10/3}\left(\frac{R_{\eta}}{R}\right)^4,\\
H_r &= \frac{2}{3}\tilde{d}_{\eta}^{-2/3} m_{\eta}^{1/3}M_P^{2/3}\left(\frac{R_{\eta}}{R}\right)^2 .
\label{rhorhr}
\end{align}
We will not concern ourselves here with the details of thermalization of either the inflaton or
AD fields \cite{ds2,am,multiple}.

The subsequent evolution of the system will depend on the value of inflaton mass.
Namely, for an inflaton mass greater than $m_c = \tilde{d}_{\eta}^{-2/3} m_{3/2}^{1/3} M_P^{2/3} \simeq \tilde{d}_{\eta}^{-2/3}10^{-5} M_P$, 
oscillations of the AD flat direction
begin after inflaton decay, i.e., $R_\Phi > R_{d\eta}$, whereas for $m_\eta < m_c$, AD oscillation begin before inflaton decay.
Remarkably, the critical mass is very close to $m_{\eta}\sim 10^{-5}M_P$ the typical mass needed to match the overall scale of density fluctuations produced by $\eta$. 

For an inflaton mass smaller than $m_c$, the oscillations of the flat direction start before the decay of $\eta$.  The scale factor at the start of AD oscillations is determined by the condition\footnote{
Technically, oscillations of $\Phi$ can not begin until $H^2 < m_{3/2}^2/3(\zeta-1)$. For $\zeta >1$,
as is required, this induces a slight delay in the onset of oscillations.  We ignore this delay in the 
analytic expressions below.}  $H \simeq 2 m_{3/2}/3$ and is given by
$R_{\Phi}/R_{\eta} = (m_{\eta}/m_{3/2})^{2/3}$.
For $m_\eta \lesssim 10^{-5} M_P$,
$R_{\Phi}/R_{d\eta} \lesssim 1$, and oscillations begin before inflaton decay so long as 
$ {\tilde d}_\eta \lesssim 1$. 
The initial amplitude of the oscillations of $\Phi$ is therefore
\begin{equation}\label{phi0_1}
|\Phi_0|= \frac{1}{\sqrt{6}}(\zeta-1)^{1/4} (m_{3/2}M_P)^{1/2}(\tilde{M}/|\lambda|)^{1/2} \sim 10^{-8}(\zeta-1)^{1/4}M_P(\tilde{M}/|\lambda|)^{1/2}.
\end{equation}
For convenience we have defined $\tilde{M}=M/M_P$. 

It is straightforward to verify that the Universe was indeed dominated by the energy density of $\eta$ at $R \le R_\Phi$, 
so that the Hubble parameter is given by Eq. (\ref{h_eta}). 
The oscillations of the flat direction carry an energy density
\begin{equation}
\rho_{\Phi}=m_{3/2}^2\Phi^2=m_{3/2}^2\Phi_0^2\left(\frac{R_{\Phi}}{R}\right)^3 \, ,
\end{equation}
where $R_\Phi$ defines the value of the scale factor when AD oscillations begin.
 At the onset of oscillations, $\rho_{\eta}/\rho_{\Phi}\sim (M_P/m_{3/2})(|\lambda|/\tilde{M})\sim 10^{15}(|\lambda|/\tilde{M})$, consistent with dominance of the inflaton $\eta$. 
 It is also straightforward to verify that the decay of the AD field occurs after the start of oscillations. 
The perturbative decay rate of the flat direction at tree level can be estimated as $\Gamma_{\Phi}\sim m_{3/2}^3/|\Phi|^2=(m_{3/2}/|\Phi_0|^2)(R_{d\Phi}/R_{\Phi})^3$, due to the large VEV of $\Phi$ \cite{AD}. At the `instant' of decay, $\Gamma_{\Phi}\sim H$, we have $R_{d\Phi}/R_{\eta}\sim 10^{9}(\zeta-1)^{1/10}\tilde{d}_{\eta}^{-2/15}(\tilde{M}/|\lambda|)^{1/5}$
indicating that $R_{d\Phi} > R_\Phi$. Note that $\Phi$ decay occurs after inflaton decay
and Eq. (\ref{rhorhr}) must be used to determine $R_{d\Phi}$.

It is convenient to write the baryon number stored in the flat direction as
\begin{equation}\label{nB}
n_{B}=\frac{n_B}{n_{\Phi}}n_{\Phi}=\frac{n_B}{n_{\Phi}}\frac{\rho_{\Phi}}{m_{3/2}}.
\end{equation}
The entropy release at reheating is $s= (4/3) (g \pi^2/30)^{1/4} \rho_{d\eta}^{3/4}= (4/3)^{7/4} (g \pi^2/30)^{1/4} \tilde{d}_{\eta}^{3}m_{\eta}^{9/2}M_P^{-3/2}$, where $g=g(T_{R})$ is the effective number of relativistic degrees of freedom at reheating. Thus the final baryon to entropy ratio can then be calculated,
\begin{equation}\label{n_s_1}
\begin{aligned}
\frac{n_B}{s} &= \frac{(3/4)^{7/4}M_P^{3/2}m_{3/2}|\Phi_0|^2}{(g \pi^2/30)^{1/4}\tilde{d}_{\eta}^3 m_{\eta}^{9/2}}\left(\frac{R_{\Phi}}{R_{d\eta}}\right)^3\left(\frac{n_B}{n_{\Phi}}\right)_{d\eta}\\
 &\simeq \frac{(3/4)^{7/4} (\zeta-1)^{1/2}}{6 (g \pi^2/30)^{1/4}} \tilde{d}_{\eta} \left(\frac{m_{\eta}}{M_P}\right)^{3/2}\frac{\tilde{M}}{|\lambda|}\left(\frac{n_B}{n_{\Phi}}\right)_{d\eta} \\
&\simeq 4\times10^{-9} g^{-1/4}(\zeta-1)^{1/2}\tilde{d}_{\eta}\left(\frac{m_{\eta}}{10^{-5}M_P}\right)^{3/2}\frac{\tilde{M}}{|\lambda|}\left(\frac{n_B}{n_{\Phi}}\right)_{d\eta} \, ,
\end{aligned}
\end{equation}
where $({n_B}/n_{\Phi})_{d\eta}$ is the ratio of baryons to $\Phi$ at the time of decay. This ratio can be approximated noting that Eq. (\ref{phi_eom}) implies the following equation for $n_B$,
\beq\label{nb_eq}
\dot{n}_B+3Hn_B=-2{\rm Im}\left(\Phi\frac{\partial V}{\partial \Phi}\right).
\eeq

After oscillations start, the right hand side of Eq. (\ref{nb_eq}) is negligible and the baryon number density per comoving volume is conserved. Therefore, $(n_B/n_{\Phi})_{d\eta}\simeq (n_{B}/n_{\Phi})_{\Phi}$. The latter ratio can be evaluated noting that at the onset of oscillations, the right hand side cannot be neglected, and is in fact of the same order as the baryon number and CP conserving terms in the potential. At this `instant' the expression (\ref{phi_min}) can be used to evaluate the right hand side of (\ref{nb_eq}), (cf. discussion  leading to (\ref{phi0_1})):
\beq
2{\rm Im}\left(\Phi\frac{\partial V}{\partial \Phi}\right)\simeq 8Am_{3/2}{\rm Im}\left(\frac{\lambda\Phi^4}{ M}\right) = \frac{A}{2}(\zeta-1) m_{3/2} H^2 \frac{M}{|\lambda|}\delta.
\eeq
Here $\delta=\sin(\arg\lambda+4\theta)\leq1$ is the $CP$ phase factor, with $\theta=\arg\Phi$. Let us ignore the phase of $\lambda$, and note that 
\beq
n_B=i(\Phi^*\dot{\Phi}-\Phi\dot{\Phi}^*)=-2|\Phi|^2\dot{\theta}.
\eeq
Before oscillations, the time dependence of $|\Phi|$ is given by (\ref{phi_min}), with $H=\frac{2}{3}t^{-1}$. Therefore, equation (\ref{nb_eq}) can be written as an equation for $\theta$. In terms of the dimensionless quantity $\tau=m_{3/2}t$, the resulting equation is
\beq
\theta''+\frac{1}{\tau}\theta'-\frac{2}{3\tau}A(\zeta-1)^{1/2} \sin(4\theta)=0.
\eeq
At the start of oscillations, $\tau=1$. For $\zeta\lesssim1+1/(36A^2)$, the friction term is very large, and the evolution is linear, $\theta'\simeq \frac{2}{3}A(\zeta-1)^{1/2}\sin(4\theta_0)$. Therefore,
\beq\label{nb_atphi}
n_{B}\simeq -\frac{4}{3}A(\zeta-1)^{1/2} m_{3/2}|\Phi_0|^2\delta_0.
\eeq
For $\zeta\gtrsim1+1/(36A^2)$, the differential equation must be solved. The initial value of $\theta$ can be restricted to $-\pi/4\leq \theta_0\leq \pi/4$ $\pmod{\pi/2}$. For small angles, the approximation $\sin(4\theta)\simeq 4\theta$ can be made. The resulting equation is solved in terms of Bessel functions. At $\tau=1$,
\beq
\theta'\simeq 2\sqrt{\frac{2}{3}A(\zeta-1)^{1/2}}\ I_{1}\left(4\sqrt{\frac{2}{3}A(\zeta-1)^{1/2}}\right) \theta_0,
\eeq
which in turn implies $n_B\sim m_{3/2}|\Phi_0|^2\delta_0$. Recall that oscillations of $\Phi$ can not begin until the effective mass of $\Phi$ becomes positive, which will affect the value of  $n_B$.  Recalling further that $n_\Phi =  m_{3/2}|\Phi_0|^2$, it is then expected that
\beq
\left(\frac{n_B}{n_{\Phi}}\right)_{\Phi}\simeq \mathcal{O}(1)\delta \leq1.
\eeq
This result is independent of the scale of the masses $m_{\eta},m_{3/2}$, and $M$. Our numerical results shown in the next section support this estimate.

Let us now assume that $m_\eta > m_c$. In this case, the decay of the inflaton occurs when $R_{d\eta}/R_{\eta}\simeq 10^6\tilde{d}_{\eta}^{-4/3}$ for $m_\eta \gtrsim 10^{-5} M_P$. The oscillations of the flat direction now start after reheating, and $R_\Phi/R_\eta =  m_\eta^{1/6} M_P^{1/3}/\tilde{d}^{1/3} m_{3/2}^{1/2}$ with $R_{\Phi}/R_{d\eta}> 1$
when  
$ {\tilde d}_\eta \gtrsim 1$.  The initial amplitude of the oscillations of the flat direction can be approximated by substituting $H_r$ evaluated at $R_\Phi$ into Eq. (\ref{phi_min}),
\begin{equation}
|\Phi_0|= \frac{1}{\sqrt{6}}(\zeta-1)^{1/4} \ m_{3/2}^{1/2}M_P^{1/2}(\tilde{M}/|\lambda|)^{1/2}\sim 10^{-8}(\zeta-1)^{1/4} M_P(\tilde{M}/|\lambda|)^{1/2}.
\end{equation}

At the onset of oscillations, the Universe is in this case dominated by the energy density of the relativistic decay products of $\eta$; $\rho_r/\rho_{\Phi}\sim (M_P/m_{\eta})^3(|\lambda|/\tilde{M}) \sim 10^{15}(|\lambda|/\tilde{M})$.
The final baryon to entropy ratio can be evaluated at the start of $\Phi$ oscillations. The entropy at $R_\Phi$ relative to the entropy released at reheating is decreased by a factor of $(R_{d\eta}/R_{\Phi})^3$. This leads to
\begin{equation}\label{n_s_2}
\begin{aligned}
\frac{n_B}{s} &= \frac{(3/4)^{7/4}M_P^{3/2}m_{3/2}|\Phi_0|^2}{4 (g \pi^2/30)^{1/4}\tilde{d}_{\eta}^3 m_{\eta}^{9/2}}\left(\frac{R_{\Phi}}{R_{d\eta}}\right)^3\left(\frac{n_B}{n_{\Phi}}\right)_{\Phi}\\
&\simeq \frac{(3/4)^{7/4}(\zeta-1)^{1/2}}{6(g \pi^2/30)^{1/4}} \left(\frac{m_{3/2}}{M_P}\right)^{1/2} \frac{\tilde{M}}{|\lambda|}\left(\frac{n_B}{n_{\Phi}}\right)_{\Phi} \\
&\simeq 4 \times10^{-9}g^{-1/4}(\zeta-1)^{1/2}  \left(\frac{m_{3/2}}{10^{-15} M_P}\right)^{1/2}\frac{\tilde{M}}{|\lambda|}\left(\frac{n_B}{n_{\Phi}}\right)_{\Phi} \, .
\end{aligned}
\end{equation}
The ratio $(n_B/n_{\Phi})_{\Phi}$ must be evaluated now in a radiation dominated Universe, $H=(2t)^{-1}$. A straightforward calculation reveals that, in this case, the baryon number density at the start of $\Phi$ oscillations is
\beq
n_{B}\simeq -\frac{A}{2}(\zeta-1) m_{3/2} \frac{MH}{|\lambda|}\delta_0 \simeq -2A(\zeta-1)^{1/2} m_{3/2}|\Phi_0|^2\delta_0
\eeq
(compare to (\ref{nb_atphi})). Therefore, $(n_B/n_{\Phi})_{\Phi}\sim\mathcal{O}(1)\delta$ is also expected in this scenario.

\section{Numerical results}

To better visualize our results, we compute
the classical time evolution for the fields in the model numerically. To obtain a form for the model appropriate for a numerical solution, the reduced Planck mass is set to one, and both the soft mass $m_{3/2}$ and the inflaton mass $m_{\eta}$ are considered within a few orders of magnitude from $M_P$. The parameters that determine the K\"ahler potential and the superpotential of the flat direction will be fixed, $\lambda/M=1$, $\zeta=2$. Since the imaginary parts of the inflaton $I$, the stabilizer $S$ and the Polonyi field $Z$ are driven to zero during inflation, we will consider vanishing initial conditions for them. For small $\Lambda$ the real part of $Z$ is small during inflation; its initial value will be taken as zero. To allow for sufficient inflation, the initial value of the inflaton $\eta$ will be taken as 20.  The only necessary constraint for the scale factor $R$ is to have a non vanishing initial value; for simplicity it will be set to unity. Table \ref{table1} summarizes these particular choices. Qualitative results are not 
affected by these choices.

\begin{table}[!h]
\centering
\begin{tabular}{|c|c|}
\hline {\bf Initial condition} & $\eta_0=20$, $\beta_0=s_0=\alpha_0=z_0=\chi_0=0$, $R_0=1$\\
\hline {\bf Parameters} & $M_P=1$, $\lambda/M=1$, $\zeta=2$, $\tilde{d}_{\eta}=1$\\
\hline
\end{tabular}
\caption{Fixed parameters for numerical evaluation.}
\label{table1}
\end{table}

The exact evolution of the classical fields and the scale factor can be found by solving the Friedmann and Klein-Gordon equations, 
\begin{equation}
H^2=\frac{1}{3}\left[K_{a\bar{b}}\dot{\Psi}^{a}\dot{\bar{\Psi}}^{\bar{b}}+V(\boldsymbol{\Psi})+\rho_{r}\right] = (\dot{R}/R)^2\ ,
\end{equation}
\begin{equation}\label{eom}
\ddot{\Psi}^{a}+3H\dot{\Psi}^{a}+\Gamma^{a}_{bc}\dot{\Psi}^b\dot{\Psi}^c+K^{a\bar{b}}\frac{\partial V}{\partial \bar{\Psi}^{\bar{b}}} = 0\ ,
\end{equation}
where the indices run from 1 to 4, $\boldsymbol{\Psi}=(I,S,\Phi,Z)$. The connection coefficients are given by
\begin{equation}\label{chriss}
\Gamma^{a}_{bc}=K^{a\bar{d}}\partial_{b}K_{c\bar{d}}\ ,
\end{equation}
and the K\"ahler metric is
\begin{equation}\label{metric}
K_{a\bar{b}}=\left(
\begin{matrix}
1 & 0 & 0 & 0\\
0 & 1- 4\xi|S|^2+\zeta|\Phi|^2 & \zeta\Phi\bar{S} & 0\\
0 & \zeta S\bar{\Phi} & 1+\zeta|S|^2 & 0\\
0 & 0 & 0 & 1-4\Lambda|Z|^2
\end{matrix}\right)\ .
\end{equation}

The energy density associated with the relativistic degrees of freedom, $\rho_r$, is assumed to correspond only to the decay products of the inflaton; $\rho_{r}=0$ for $t<t_{d\eta}$. It is modeled as
\beq
\rho_r = \rho(\eta)_{d\eta}\left(\frac{R_{d\eta}}{R}\right)^4,
\eeq
where the energy density $\rho_{d\eta}$ and the scale factor $R_{d\eta}$ at the time of decay are determined from continuity with the numerical solution. The evolution of the AD field is followed until its decay, at $t\simeq \Gamma_{\Phi}^{-1}$.

Two numerical solutions are presented below. The parameters chosen for each one correspond to the following,
\begin{itemize}
\item[(a)] $m_{\eta}=10^{-1}$, $\mu^2=\sqrt{3}\times10^{-4}$, $\Lambda^2=10^{-1}$, $\xi=1$, $\phi_0=1/10$, $\gamma_0=1/20$,
\item[(b)] $m_{\eta}=10^{-2}$, $\mu^2=10^{-5}$, $\Lambda^2=10^{-2}$, $\xi=10$, $\phi_0=1/20$, $\gamma_0=1/10$.
\end{itemize}
Recall that $m_{3/2}=\mu^2/\sqrt{3}$. Case (a) realizes the scenario in which the oscillations of the AD field occur after reheating, $R_{d\eta}<R_{\Phi}$ or $m_\eta > m_c$; in turn (b) corresponds to the scenario in which the oscillations of $\Phi$ start before the decay of the inflaton, $R_{d\eta}>R_{\Phi}$ 
or $m_\eta < m_c$. In both cases, the flat direction decays after the onset of oscillations of $\Phi$ and after reheating, $R_{d\Phi}>R_{d\eta},R_{\Phi}$. The qualitative evolutionary features found here are expected to persist for a more 
realistic separation of masses.

\begin{figure}[ht!]
\centering
 	\scalebox{.7}{\includegraphics{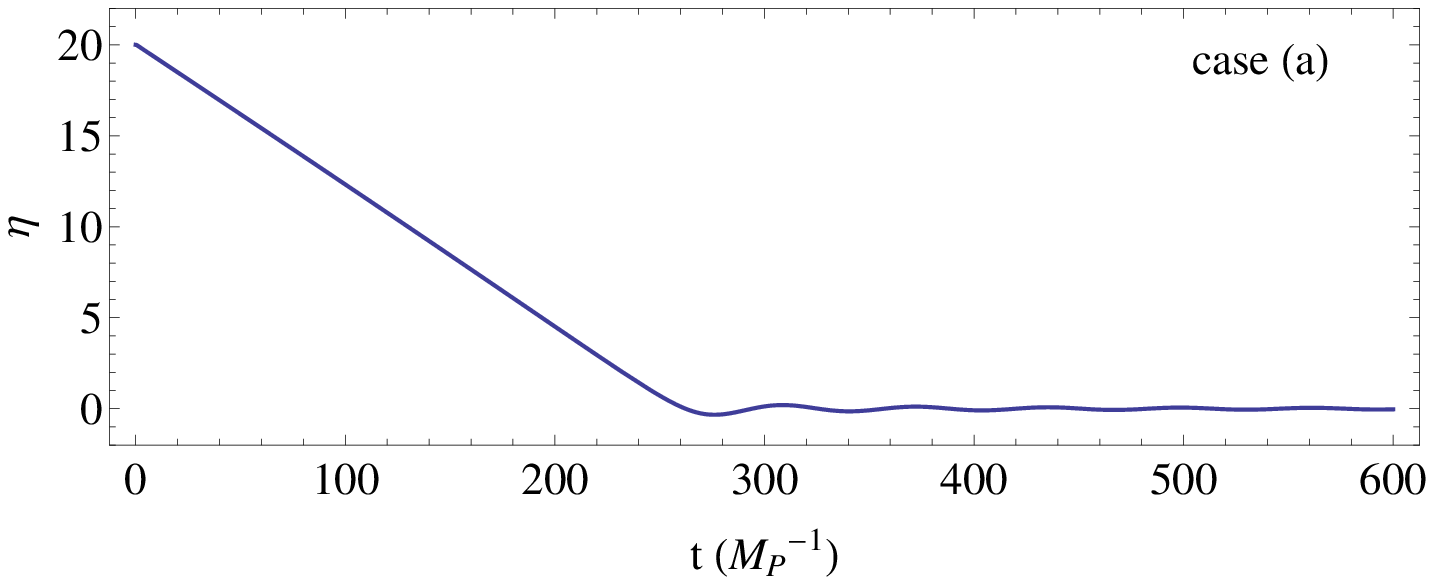}}\\
	\scalebox{.7}{\hspace{-2em}\includegraphics{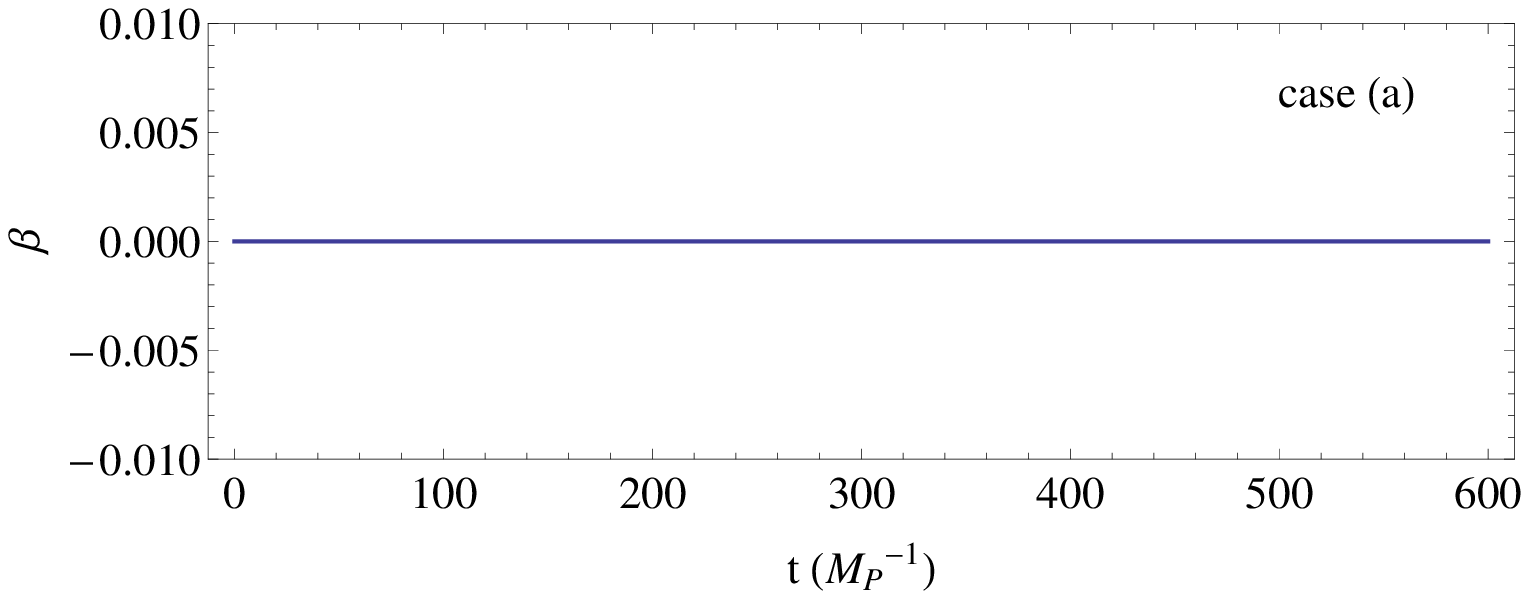}}
	\caption{Evolution of the real and imaginary parts of the inflaton $I$, in case (a). The real part $\eta$ rolls slowly towards the minimum during inflation, while the imaginary part $\beta$ is driven to zero and remains vanishingly small. The decay of $\eta$ occurs at $t\simeq 10^3 M_P^{-1}$.}
	\label{eta1}
\end{figure}

\begin{figure}[ht!]
\centering
 	\scalebox{.7}{\includegraphics{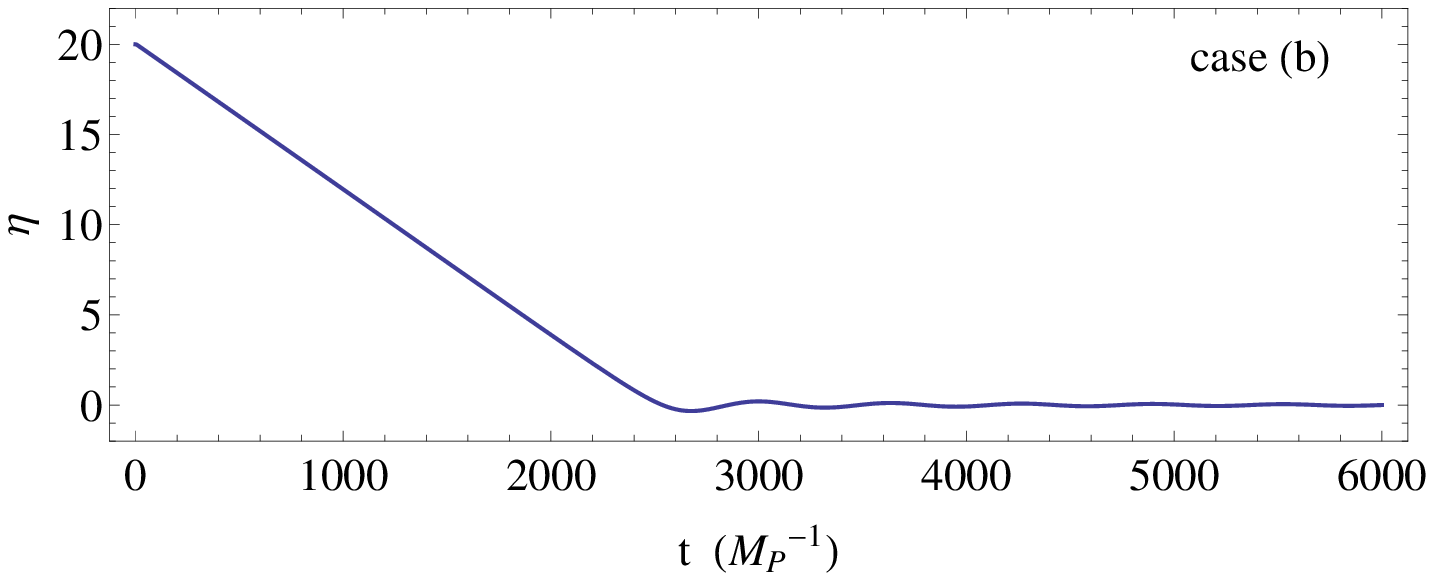}}\\
	\scalebox{.7}{\hspace{-1.2em}\includegraphics{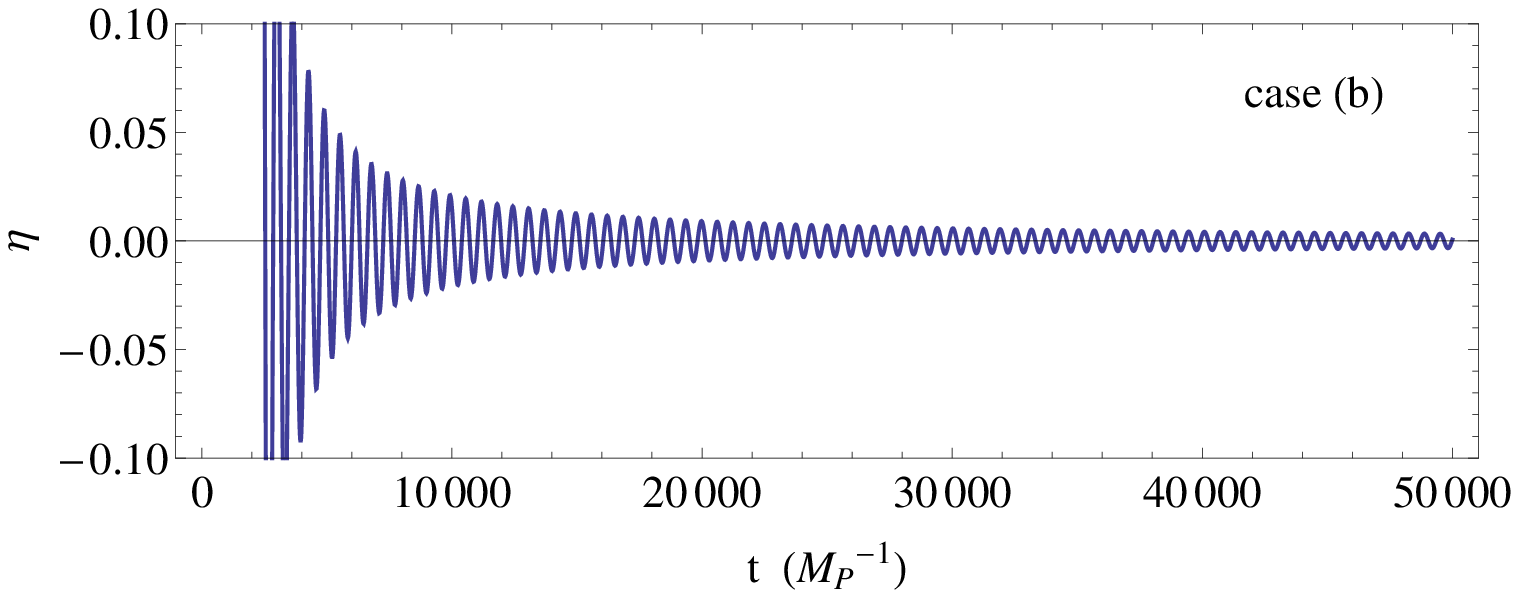}}
\caption{Evolution of the inflaton $\eta$, in case (b). The first graph shows the slow roll during inflation; the second shows the matter-like oscillations of $\eta$ at the end of inflation. The inflaton decays at $t\simeq 10^6 M_P^{-1}$.}
\label{eta2}
\end{figure}

\begin{figure}[!h]
\centering
 	\scalebox{.7}{\includegraphics{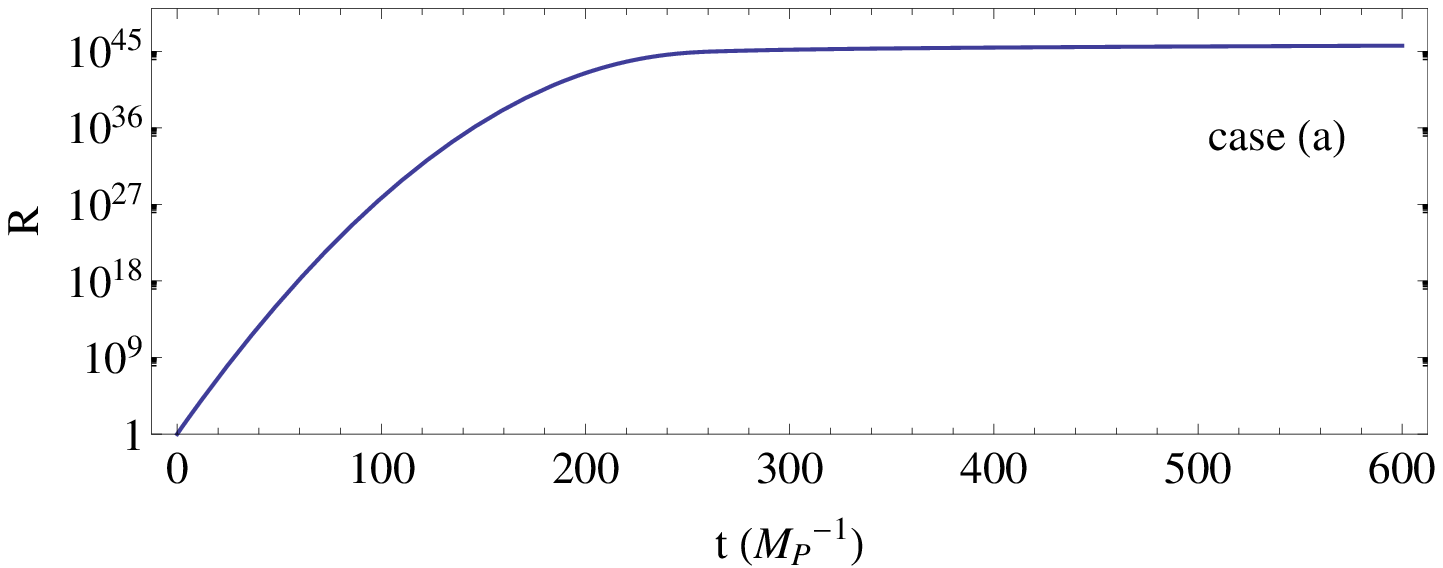}}\\[5pt]
	\scalebox{.7}{\includegraphics{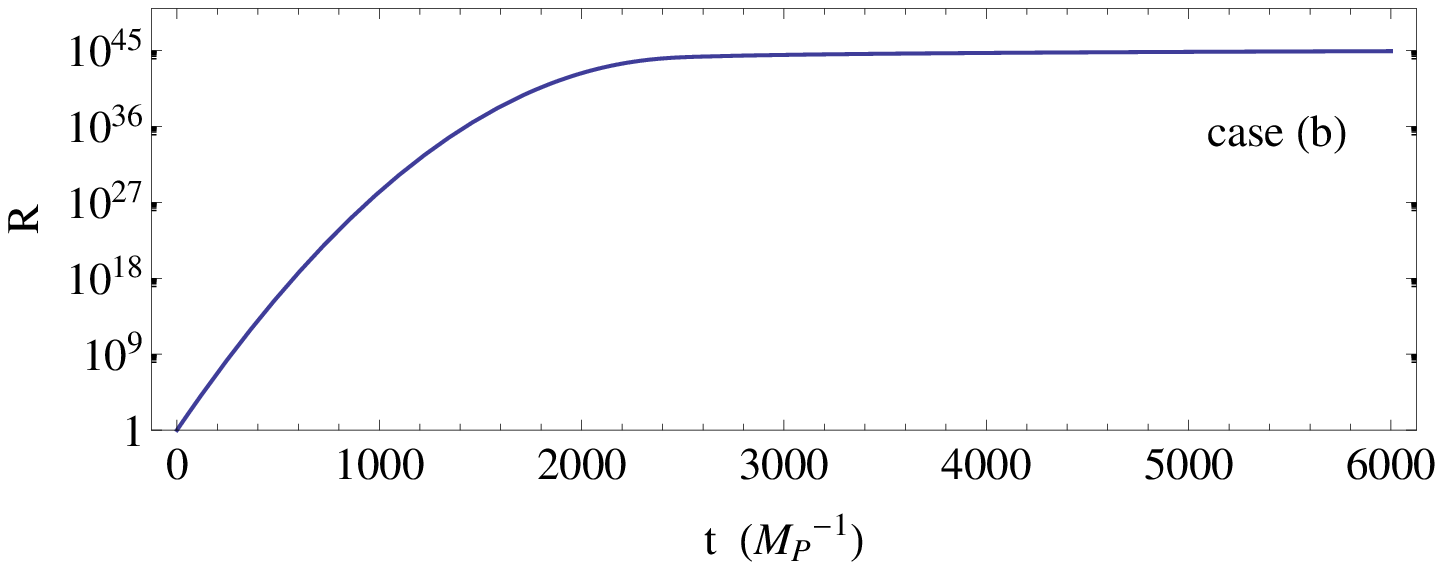}}
	\caption{Growth in time of the scale parameter $R$ in cases (a) and (b). In both cases, $R$ grows during inflation by a factor of $\sim e^{100}$. Notice the difference in the time-scales between the two cases.}
\label{R}
\end{figure}

Figure \ref{eta1} shows the evolution of $I$ in case (a). As is expected, the real part $\eta$ rolls slowly towards the minimum, realizing the chaotic inflation scenario. Perturbations to this motion due to the presence of $Z$ and $\Phi$ are not evident. The imaginary part $\beta$ is also unperturbed, and remains small during and after inflation. As a consequence, for case (b), $\beta$ was set to zero and  Figure \ref{eta2} only shows the evolution of the inflaton $\eta$.  Here the two panels correspond to different time scales.  In the lower panel, the scale is blown up so as to better see the damped matter-like coherent oscillations of $\eta$ at late times. Figures \ref{R}  demonstrate the exponential growth of the scale factor $R$ during the inflationary stage. For the initial condition of $\eta$ considered, $R$ grows by approximately 100 e-folds.

\begin{figure}[!h]
\centering
	\scalebox{.7}{\hspace{-2em}\includegraphics{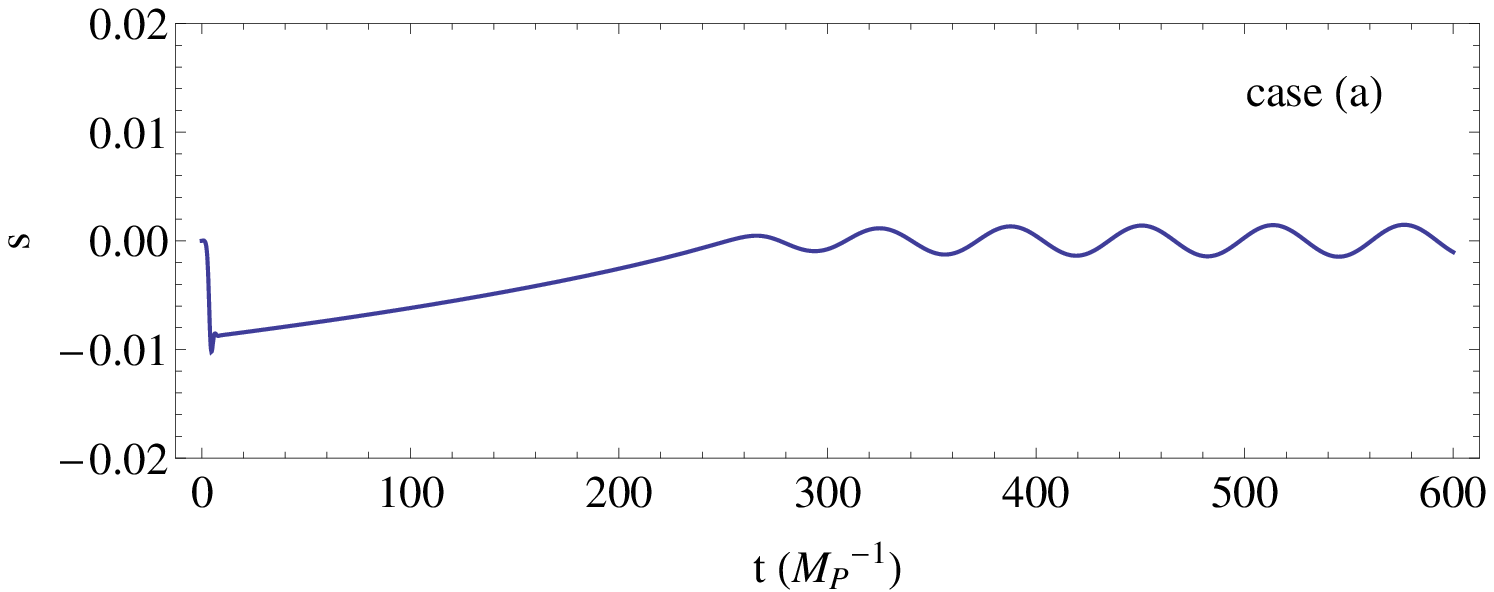}}\\[5pt]
 	\scalebox{.7}{\hspace{-2em}\includegraphics{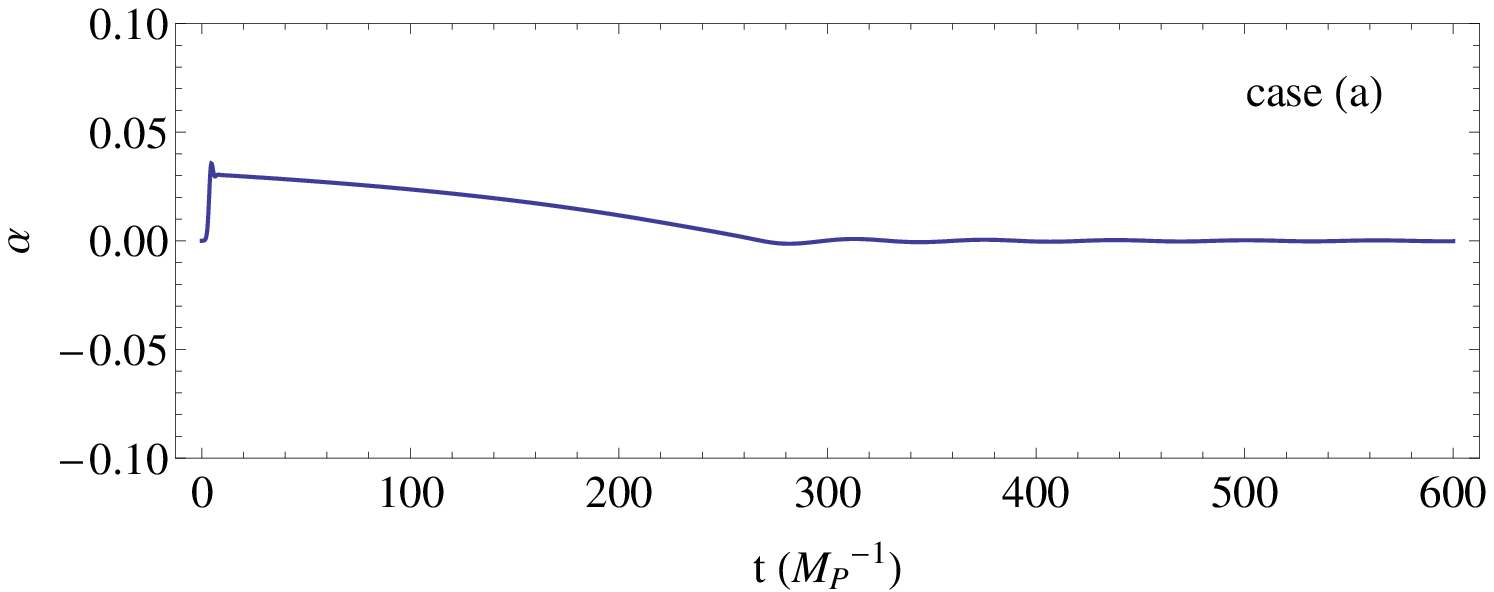}}
	\caption{Evolution of the real and imaginary parts of the stabilizer $S$, in case (a). Both the real part $s$ and the imaginary part $\alpha$ remain small during inflation. At the onset of $\eta$ oscillations, $S$ oscillates about the origin.}
\label{s1}
\end{figure}

\begin{figure}[!h]
\centering
	\scalebox{.7}{\hspace{-2.2em}\includegraphics{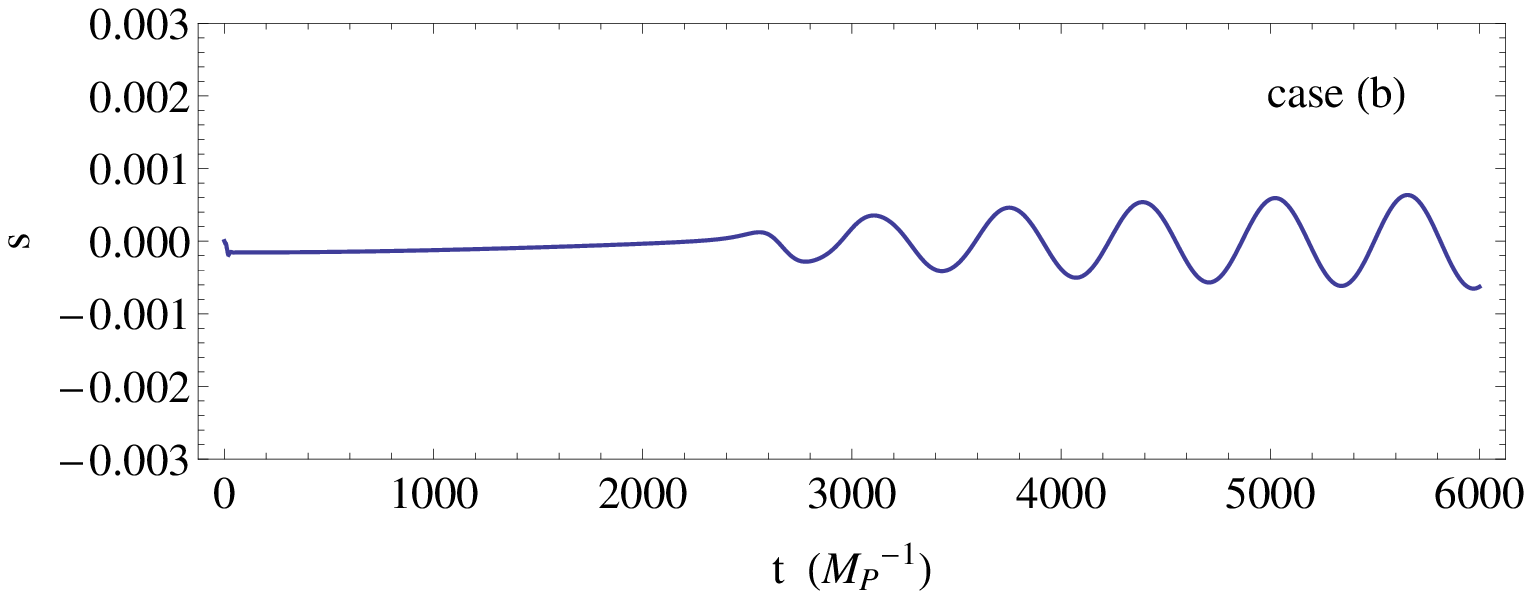}}\\[10pt]
 	\scalebox{.7}{\hspace{-2em}\includegraphics{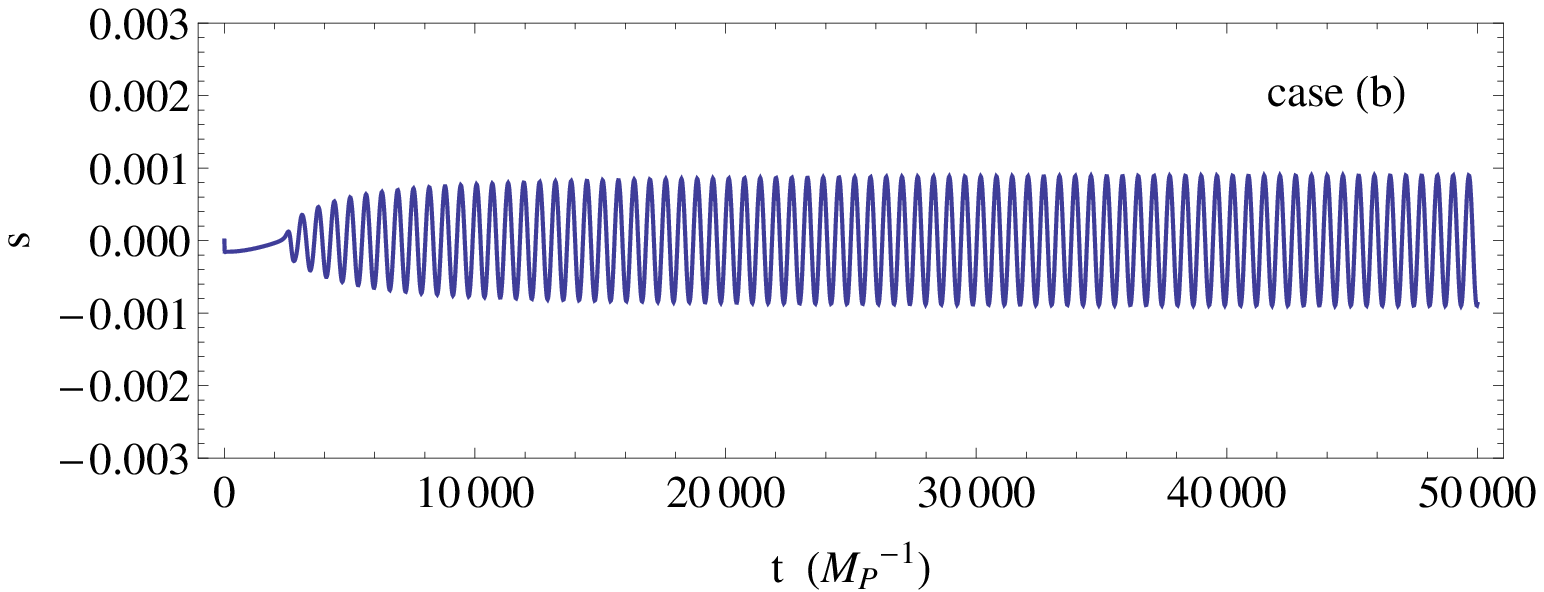}}
\caption{Evolution of the real part of the stabilizer, $s$, in case (b). During the inflationary phase, $s$ remains small. When inflation ends, $s$ oscillates about the minimum $s=0$.}
\label{s2}
\end{figure}

In Figure \ref{s1}, we show the time dependence of the stabilizer field $S$  for case (a). It is worth emphasizing that $S$ remains close to zero during inflation, even when in this scenario the stabilizing parameter $\xi$ was set to one. When inflation ends, the inflaton $\eta$ and the real part of $S$ oscillate about the origin. At this minimum, the masses are $m_s^2=m_{\alpha}^2=m_{\beta}^2=m_{\eta}^2$. Thus, $s$ oscillates with the same frequency as $\eta$, but with a much smaller amplitude. The imaginary part $\alpha$ is relatively unperturbed, and was therefore set to zero in case (b). Figure \ref{s2} shows the evolution of $s$ in the second case. Notice that the amplitude of oscillations of $s$ is more than one order of magnitude smaller than those of $\eta$. 

\begin{figure}[!h]
\centering
	\scalebox{.7}{\hspace{-1.6em}\includegraphics{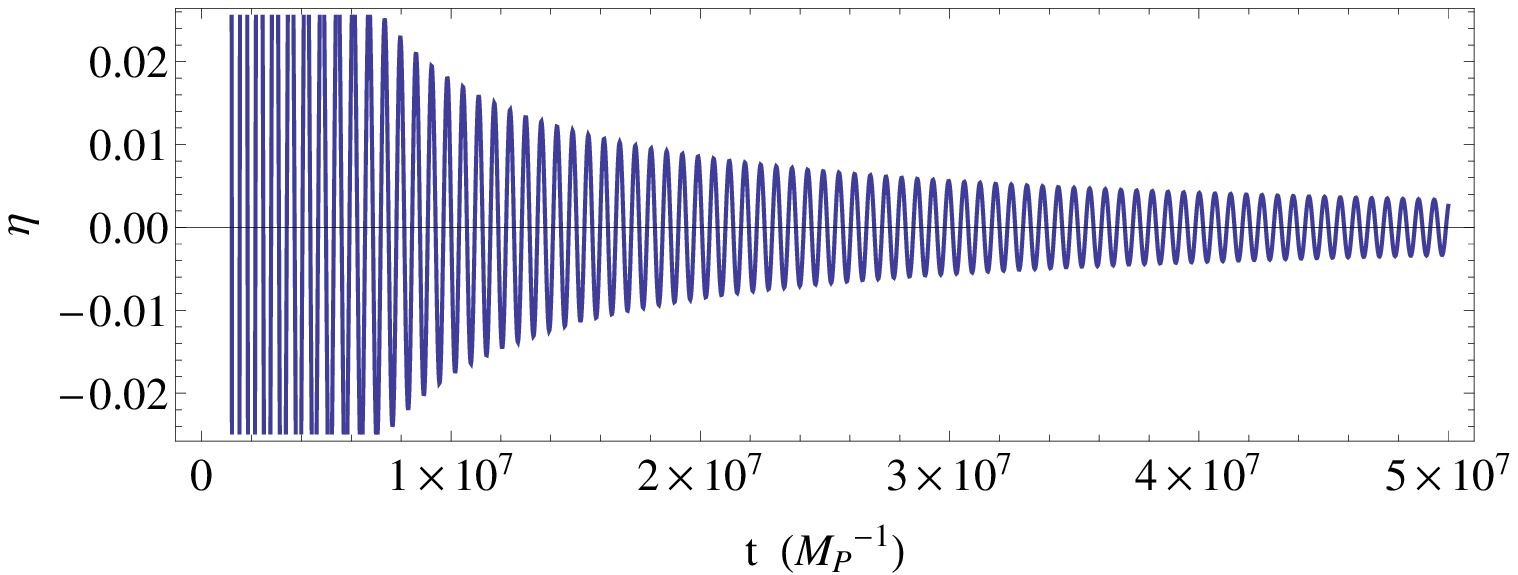}}\\[5pt]
 	\scalebox{.7}{\hspace{-4em}\includegraphics{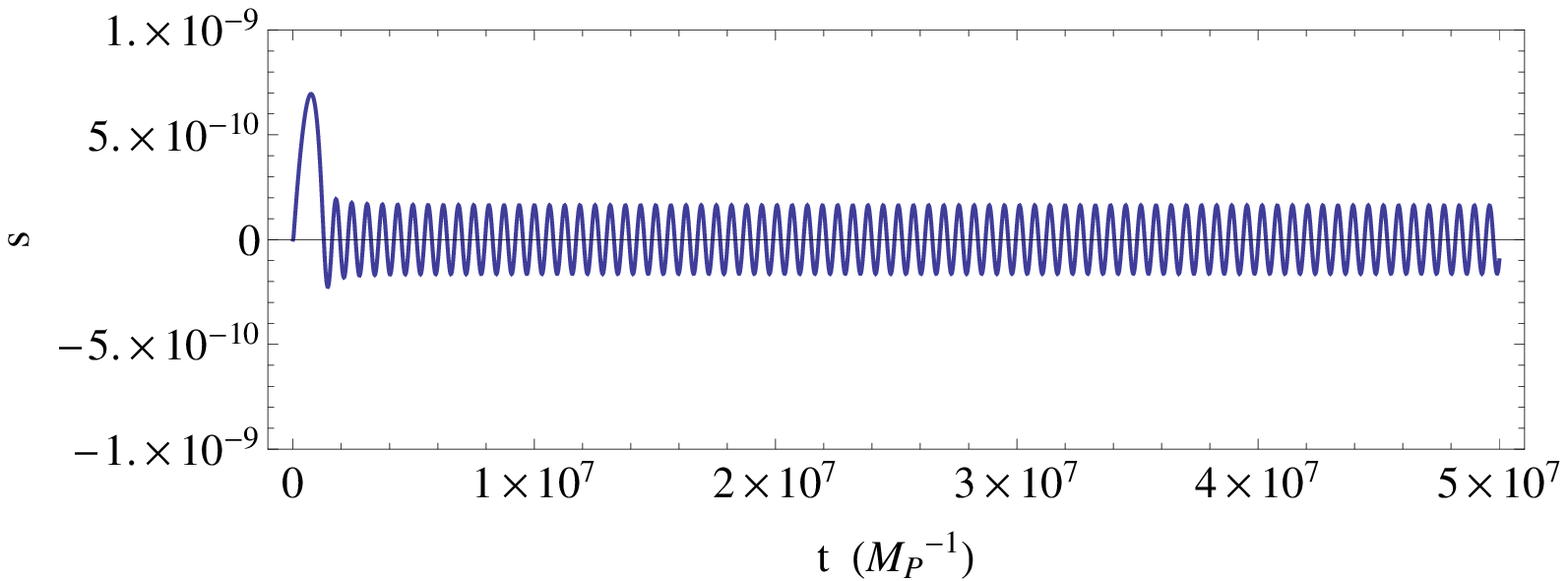}}
\caption{Numerical solution of equations (\ref{driven1}), (\ref{driven2}) for $m_{\eta}=10^{-5}$, $m_{3/2}=10^{-15}$ (in Planck units). The oscillations of $s$ are driven by the amplitude of $\eta$, but suppressed by nine orders of magnitude }
\label{driven}
\end{figure}

The oscillatory behavior of $s$ after inflation can be traced to the non-vanishing Polonyi superpotential, which induces a coupling between the inflaton and the stabilizer,
\beq
\begin{aligned}
K^{S\bar{S}}D_SW\bar{D}_{\bar{S}}\bar{W}-3W\bar{W} &= K_{S}W_{\text{\cancel{susy}}}\partial_{\bar{S}}\bar{W}_{\rm inf}+\partial_{S}W_{\rm inf}K_{\bar{S}}\bar{W}_{\text{\cancel{susy}}} + \cdots\\
&\simeq \mu^2\nu \left[Sf(I)+\bar{S}\bar{f}(\bar{I})\right] + \cdots
\end{aligned}
\eeq
For the present analysis, $f(I)=m_{\eta}I$. When the oscillations of $\eta$ start, the equations of motion are
\begin{align}
\ddot{\eta}+3H\dot{\eta}+m_{\eta}^2\eta&=2m_{3/2}m_{\eta}s, \label{driven1}\\
\ddot{s}+3H\dot{s}+m_{\eta}^2s&=2m_{3/2}m_{\eta}\eta. \label{driven2}
\end{align}
In our numerical model, $m_{\eta}$ and $m_{3/2}$ are artificially close, and the stabilizer $s$ is driven by the inflaton to relatively large values. In a more realistic scenario, $m_{3/2}\ll m_{\eta}$, and the right hand side of (\ref{driven1}), (\ref{driven2}) is almost negligible, leading to $s\ll \eta$. Figure \ref{driven} exemplifies this.

\begin{figure}[!h]
\centering
	\scalebox{.7}{\hspace{-1em}\includegraphics{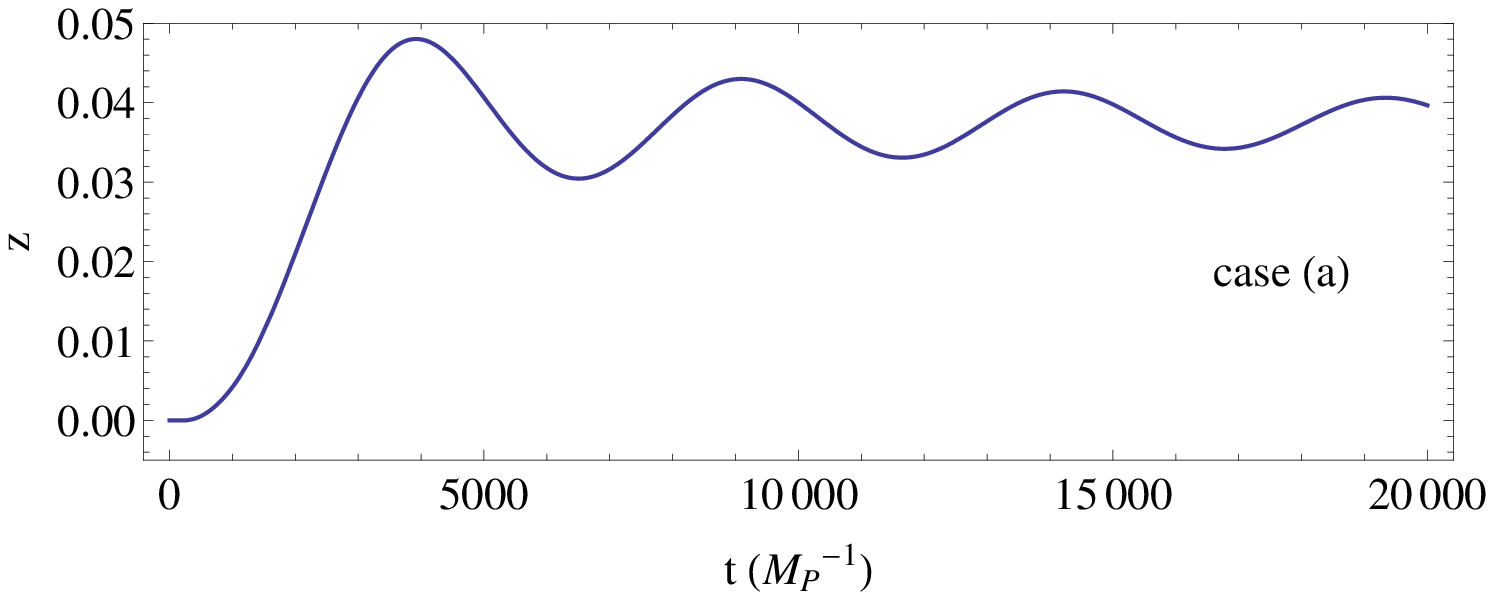}}\\[5pt]
	\scalebox{.7}{\hspace{-2.2em}\includegraphics{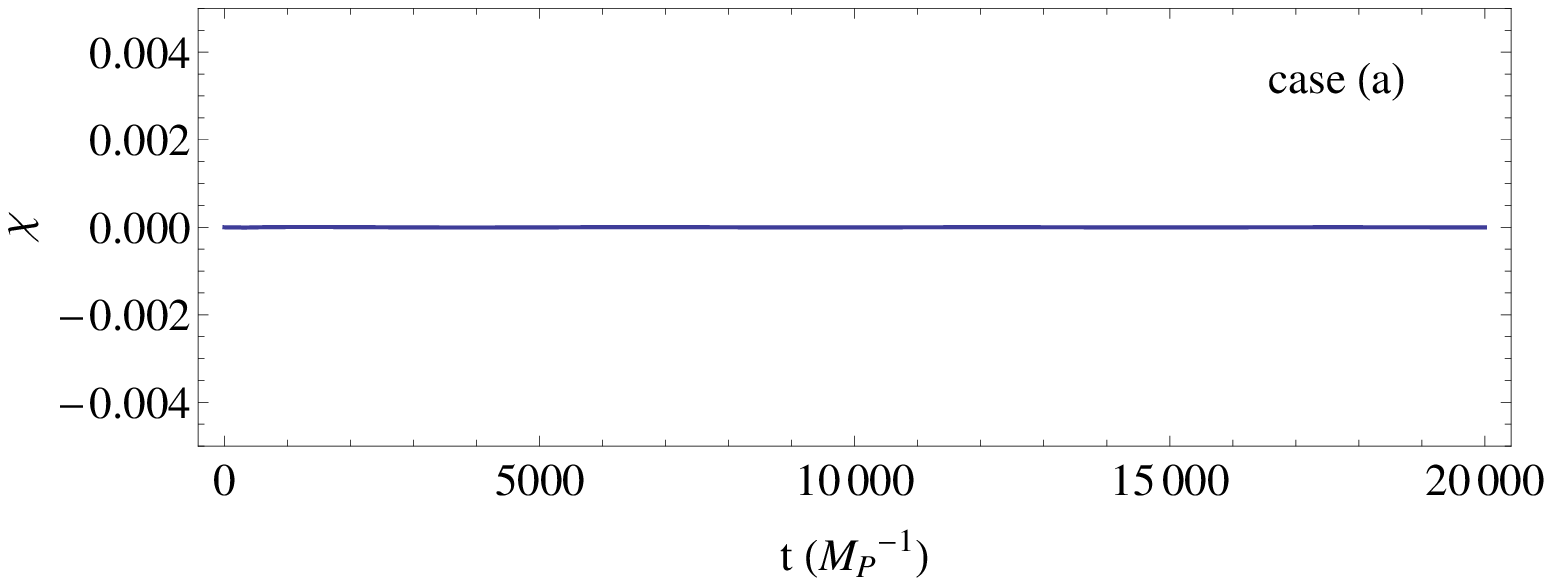}}
	\caption{Evolution of the real and imaginary parts of the Polonyi field $Z$, in case (a). The imaginary part $\chi$ remains at zero, while the real part $z$ interpolates between the minimum during inflation (\ref{zmininf}) and the true minimum (\ref{z_min}).}
\label{z1}
\end{figure}

\begin{figure}[!h]
\centering
	\scalebox{0.7}{\hspace{-1.8em}\includegraphics{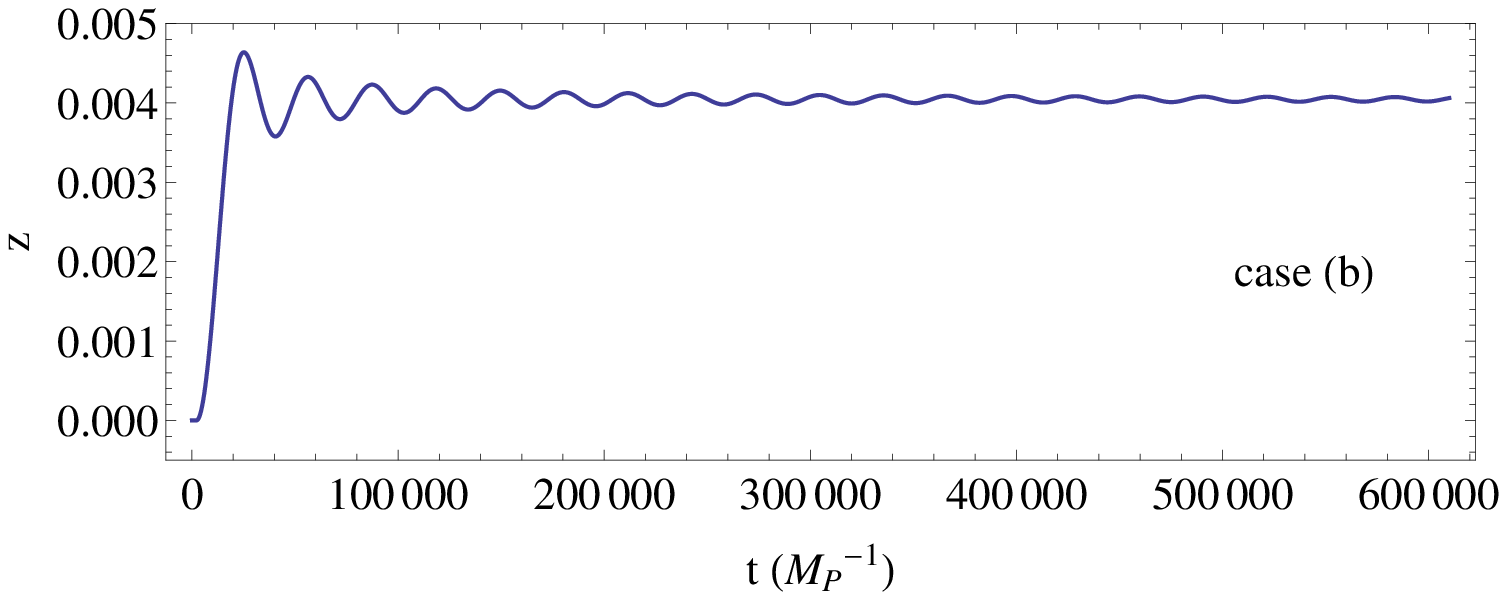}}
	\caption{Time dependence of the real part of the Polonyi field, $z$, in case (b). The small value of $\Lambda$ stabilizes $z$, damping the oscillations about its minimum.}
\label{z2}
\end{figure}

The time evolution of the Polonyi field $Z$ in case (a) is shown in Figure \ref{z1}. During the inflationary phase, the real part $z$ is driven to the instantaneous minimum (\ref{zmininf2}), which is very close to zero. The imaginary part $\chi$ vanishes. After inflation ends, the Hubble parameter starts decreasing and $z$ evolves quickly to the supersymmetry breaking minimum (\ref{z_min}). Indeed we see some form of adiabatic relaxation \cite{adrel} taking place as the amplitude of oscillations of $z$ is far smaller than the total displacement of the minimum during inflation.  Figure \ref{z2}, for case (b) in which $\chi=0$, shows explicitly how the presence of the stabilizing term $|Z|^2/\Lambda^2$ in the K\"ahler potential (\ref{okklt}) damps the oscillation of $z$ about its minimum. This is necessary to address the cosmological moduli problem.

\begin{figure}[!h]
\centering
 	\scalebox{0.7}{\hspace{-0.5em}\includegraphics{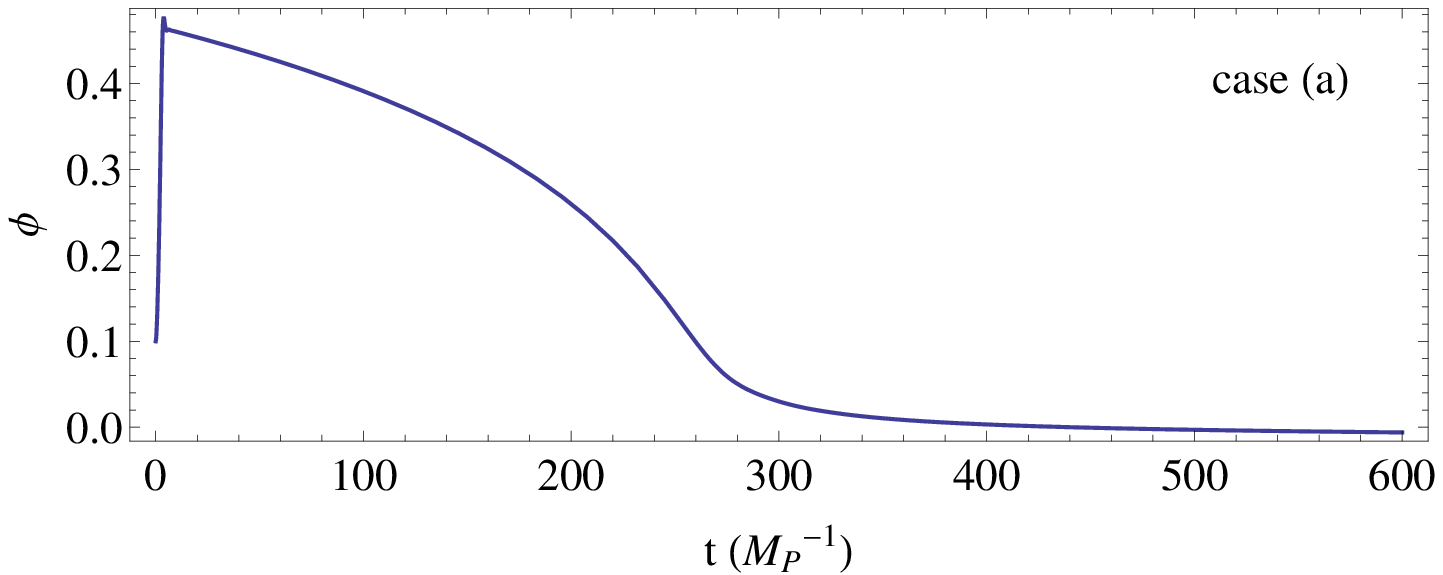}}\\[10pt]
	\scalebox{0.7}{\hspace{-2.5em}\includegraphics{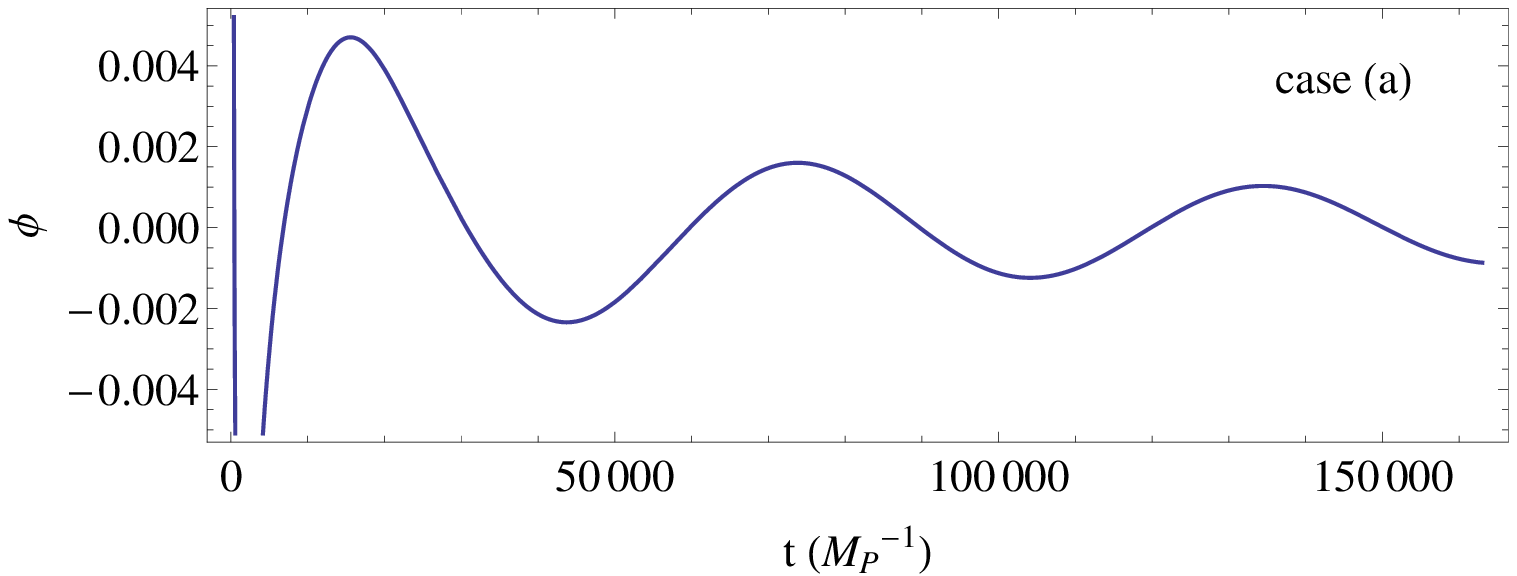}}
\caption{Time dependence of the real part of the AD field $\Phi$, in case (a). During inflation, $\phi$ tracks the instantaneous minimum (\ref{phi_min}). For $t\gtrsim 7500 M_P^{-1}$, it oscillates about $\phi=0$.}
\label{ADr1}
\end{figure}

\begin{figure}[!h]
\centering
 	\scalebox{.7}{\hspace{-0.8em}\includegraphics{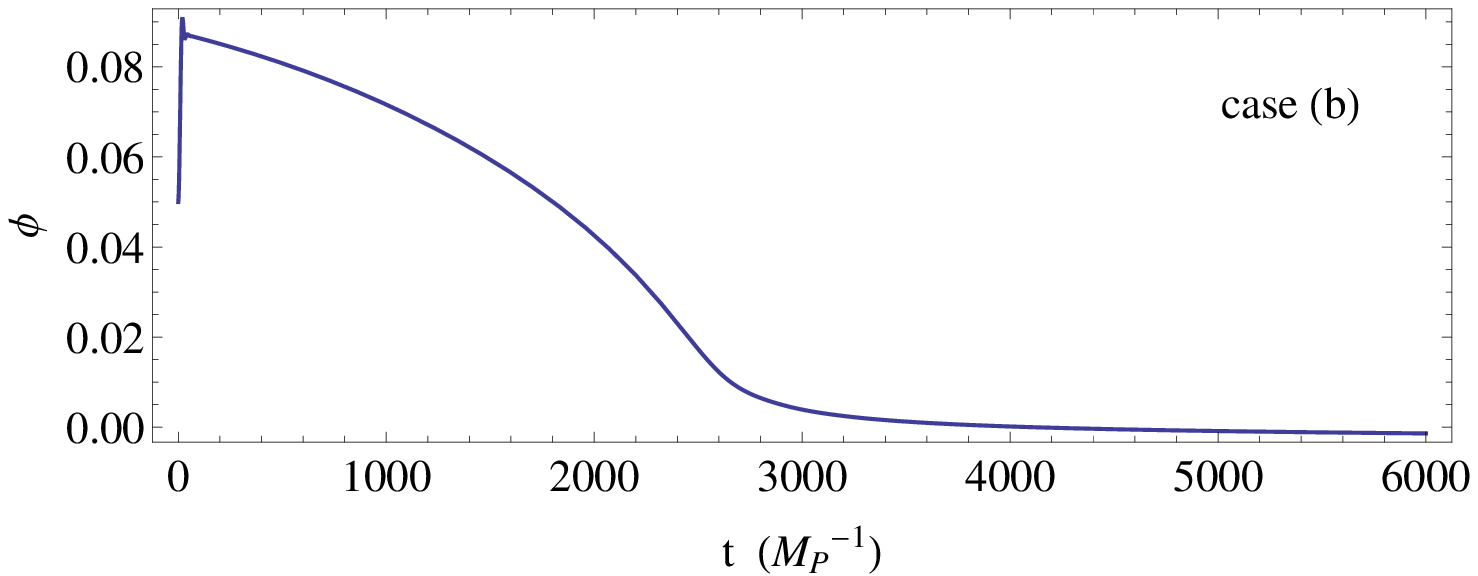}}\\[10pt]
	\scalebox{.7}{\hspace{-1.9em}\includegraphics{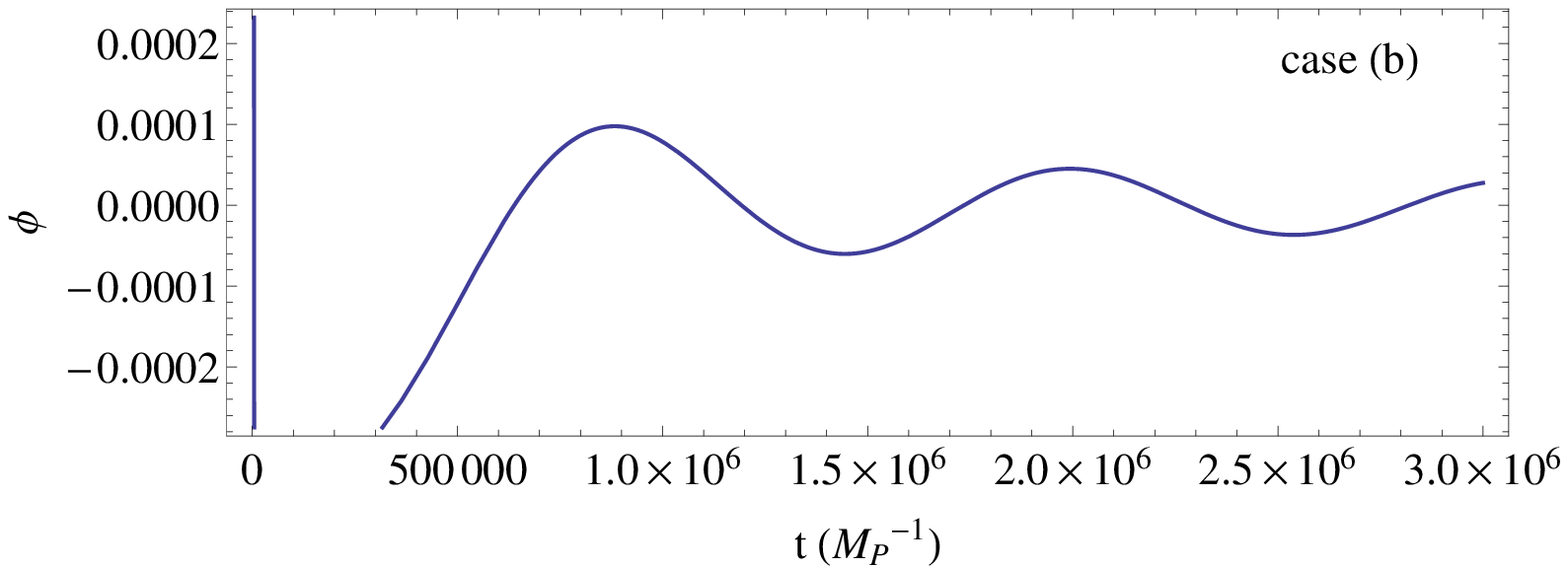}}
	\caption{Time evolution of the real part of the AD field $\Phi$, in case (b), during inflation, tracking the instantaneous minimum, and during $\Phi$ oscillations about the origin ($t\gtrsim 1.7\times10^5 M_P^{-1}$).}
\label{ADr2}
\end{figure}

\begin{figure}[!h]
\centering
	\scalebox{.7}{\hspace{-1em}\includegraphics{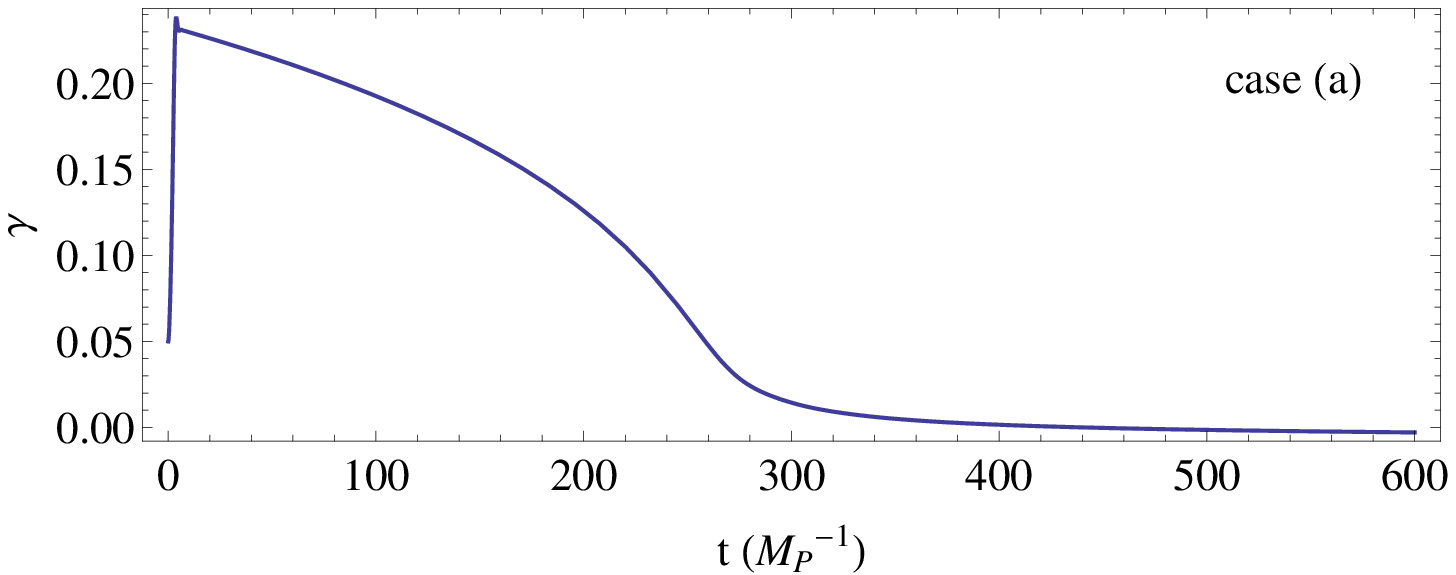}}\\[10pt]
 	\scalebox{.7}{\hspace{-2.5em}\includegraphics{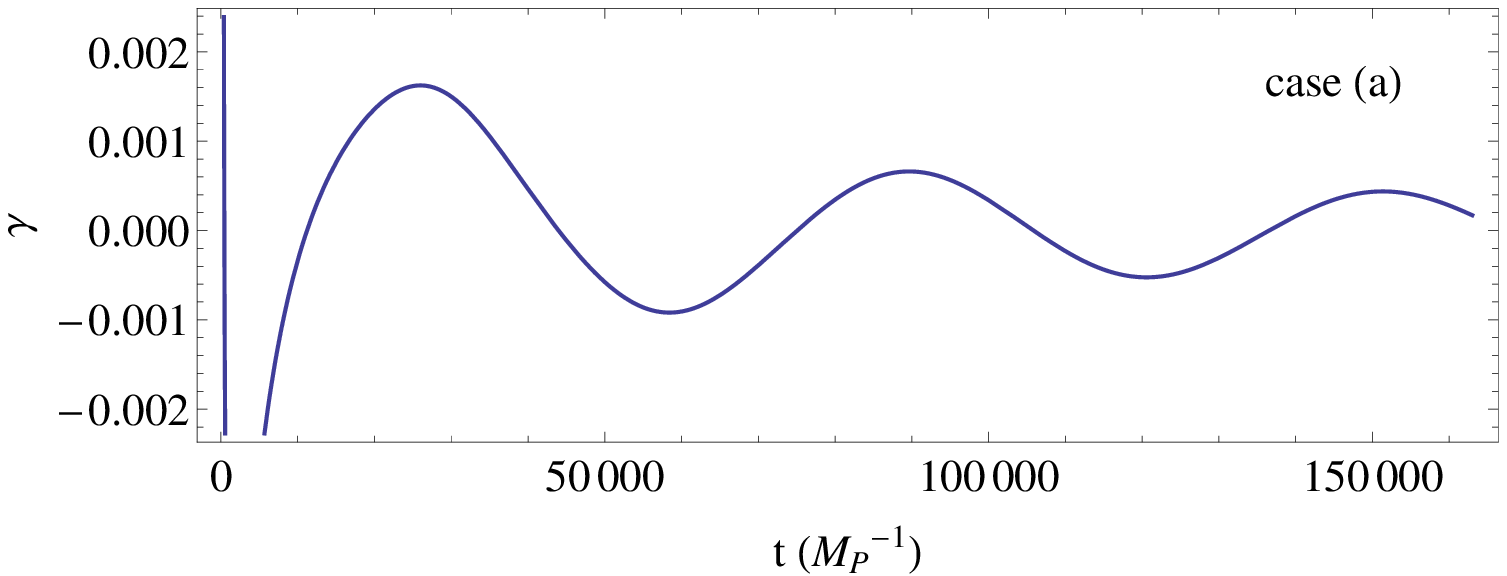}}
\caption{Time dependence of the imaginary part of the AD field $\Phi$, in case (a). During inflation, $\gamma$ tracks the instantaneous minimum (\ref{phi_min}). For $H\leq m_{3/2}$, it oscillates about $\gamma=0$, out of phase with $\phi$.}
\label{ADi1}
\end{figure}

\begin{figure}[!h]
\centering
	\scalebox{.7}{\hspace{-0.8em}\includegraphics{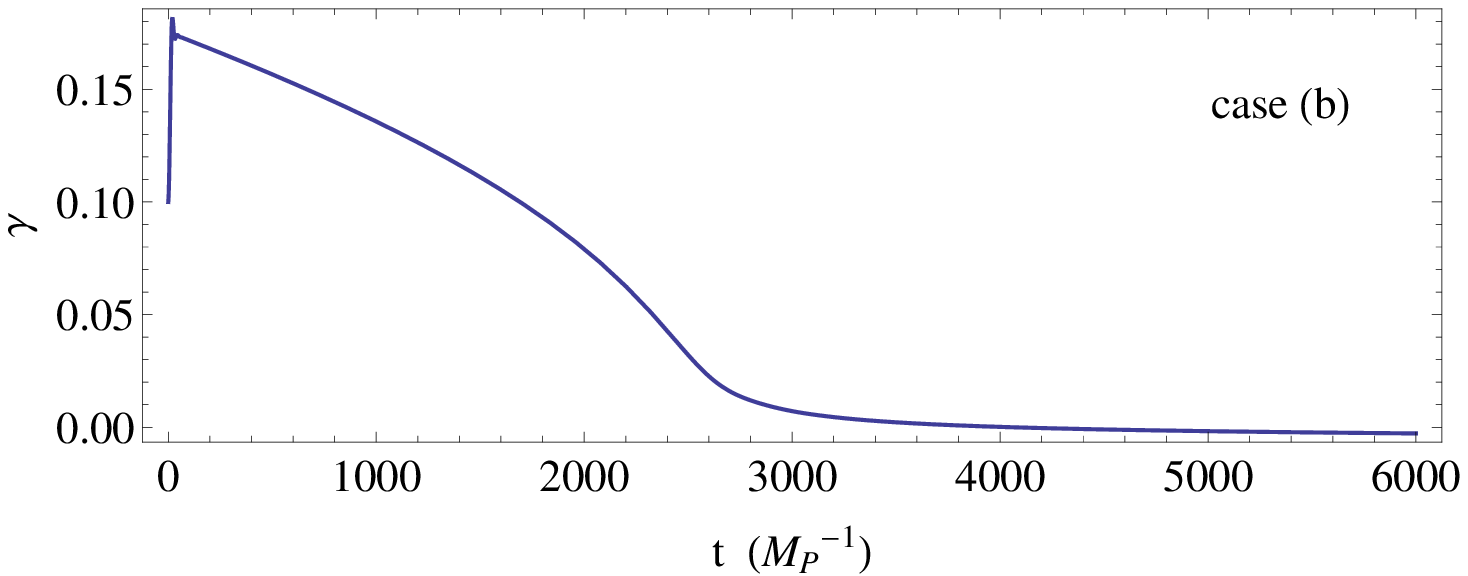}}\\[15pt]
 	\scalebox{.7}{\hspace{-1.8em}\includegraphics{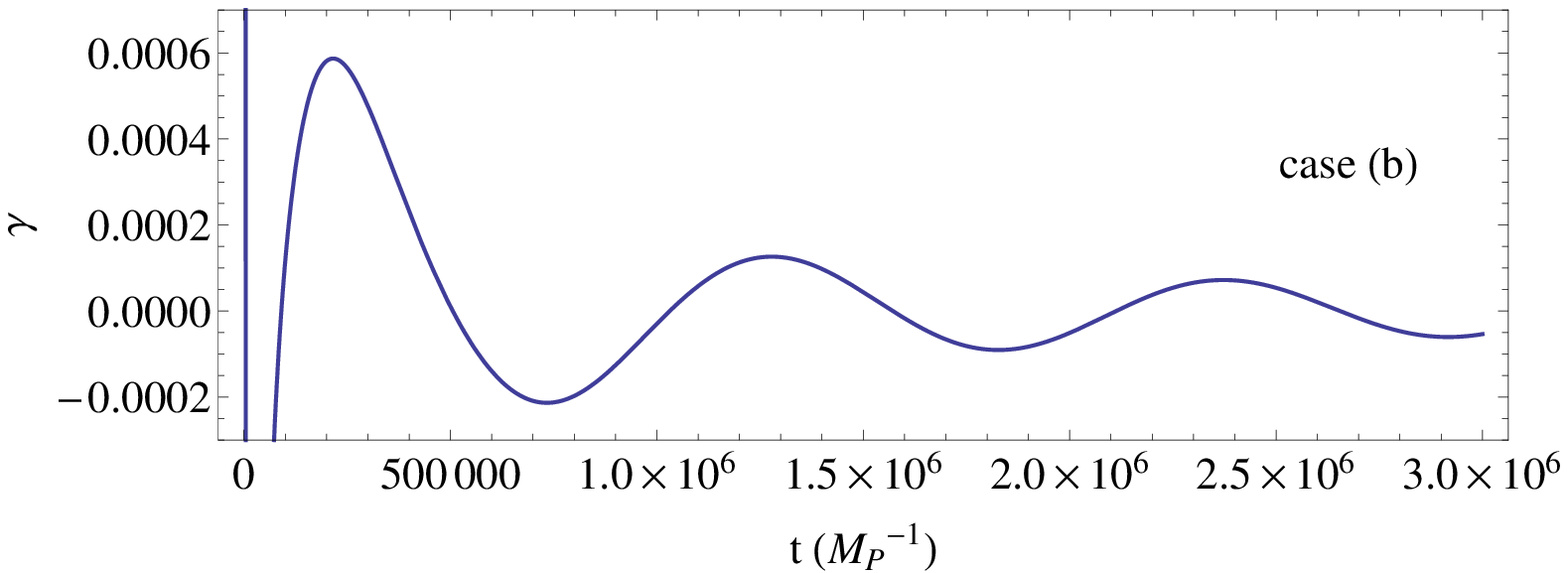}}
	\caption{Time evolution of the imaginary part of the AD field $\Phi$, in case (b), during inflation and during $\Phi$ oscillations. Note that $\gamma$ oscillates out of phase with $\phi$.}
\label{ADi2}
\end{figure}

\begin{figure}[!h]
\centering
 	\scalebox{.7}{\includegraphics{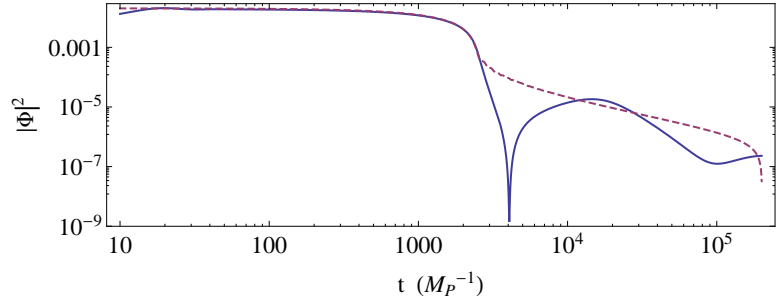}}
	\caption{Tracking of the instantaneous minimum (\ref{phi_min}), for (b). The dashed purple curve corresponds to the minimum (\ref{phi_min}). Notice that $|\Phi|$ tracks the minimum closely until the end of inflation, when it overshoots it and starts oscillations with decreasing amplitude.}
\label{trck}
\end{figure}

Figures \ref{ADr1} and \ref{ADr2} show the time dependence of the real part of $\Phi$, for both scenarios. Figures \ref{ADi1} and \ref{ADi2} demonstrate this dependence for the imaginary part. 
In each of the four figures the lower panel shows the evolution over an expanded time scale and an
enhanced amplitude. The random $CP$ violating phase due to the de Sitter fluctuation of $\Phi$ appears as the initial values of these fields which we have taken as $\phi_0 =1/10, \gamma_0 = 1/20$ for case (a) and  $\phi_0 = 1/20, \gamma_0 = 1/10$ for case (b). During inflation, $\Phi$ is driven to the instantaneous minimum. In (a), substitution of $\eta\simeq 20$ in (\ref{phi_min}) predicts an initial value of $\phi\simeq 0.57$, $\gamma\simeq 0.28$; this approximation is not precise since $0.1\lesssim|\Phi_0|\lesssim 1$. Using the full potential (\ref{AD_potential}), the position of the minimum is found to be $\phi\simeq 0.46$, $\gamma\simeq 0.23$. Therefore, the AD field evolves rapidly to the instantaneous minimum. Figure \ref{trck} shows that this evolution is exponentially fast. $\Phi$ tracks closely the instantaneous minimum during inflation, until the inflaton starts its oscillations. In Figure \ref{trck}, the sudden change in behavior of the Hubble parameter can be appreciated. $\Phi$ then overshoots the minimum, and starts damped oscillations about this minimum.

\begin{figure}[!h]
\centering
	\scalebox{0.6}{\includegraphics{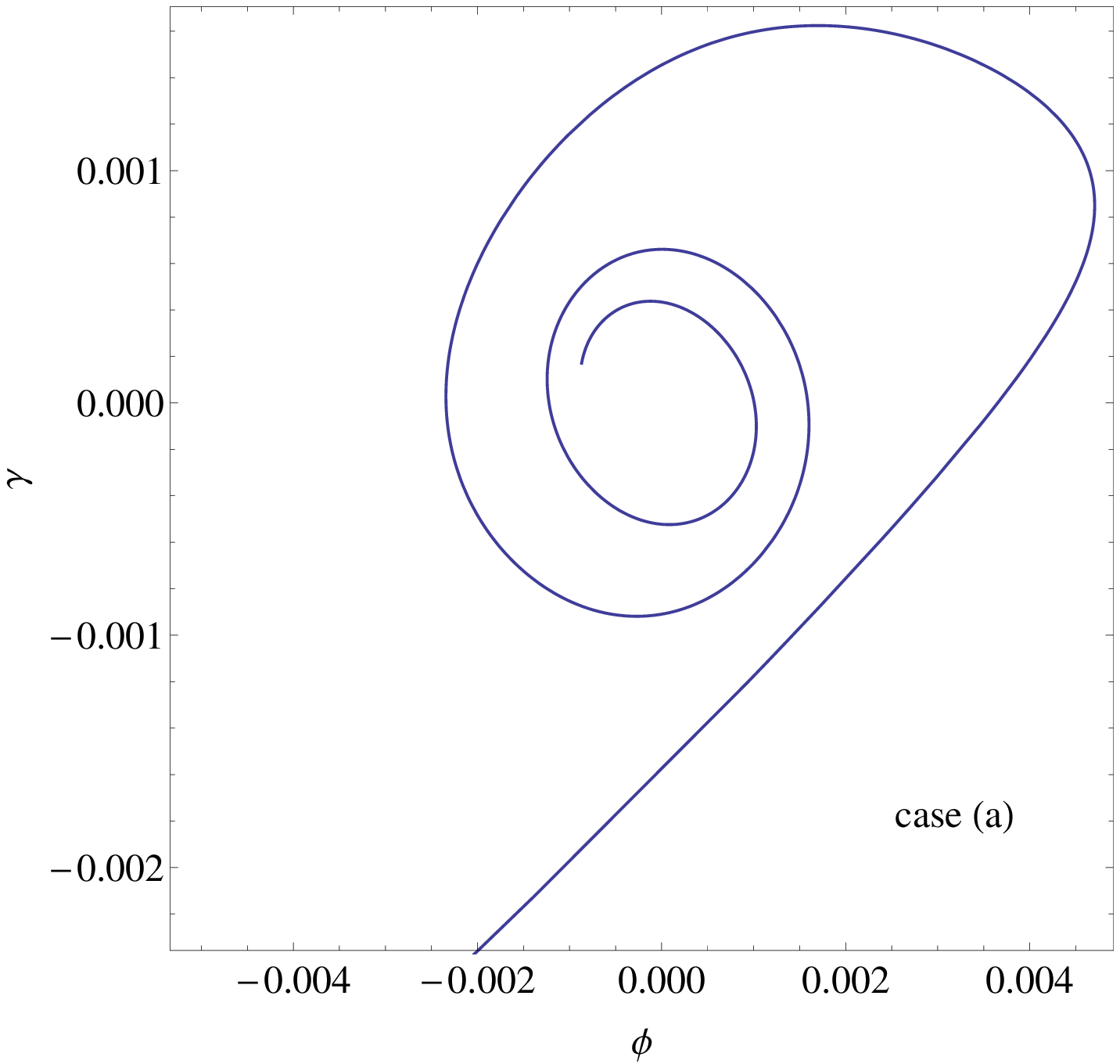}}\\[10pt]
	\scalebox{0.7}{\hspace{-1.5em}\includegraphics{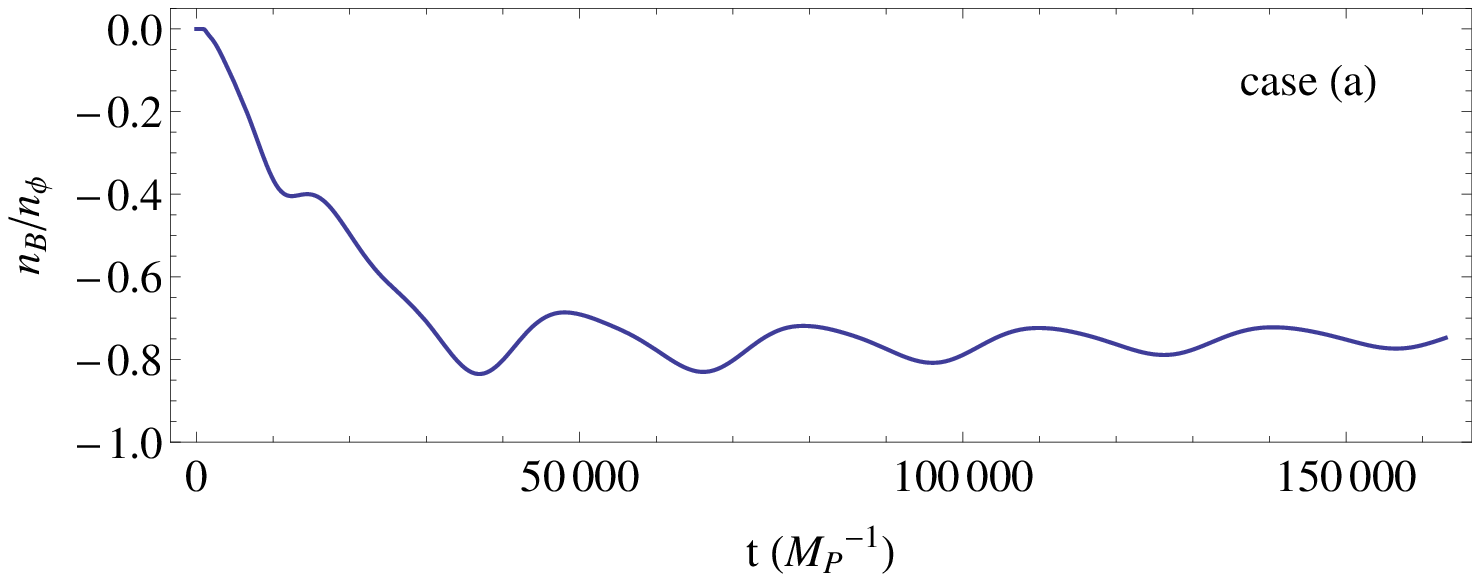}}
	\caption{Evolution of $\Phi$ in the complex plane for $H\leq m_{3/2}$, and baryon asymmetry in case (a). The baryon number is stored in the spiral motion. After the onset of $\Phi$ oscillations, $n_B/n_{\Phi}$ evolves towards a constant value of order one. $\Phi$ eventually decays at $t\simeq 1.64\times10^{5} M_P^{-1}$.}
\label{ADs1}
\end{figure}

\begin{figure}[!h]
\centering
	\scalebox{0.6}{\includegraphics{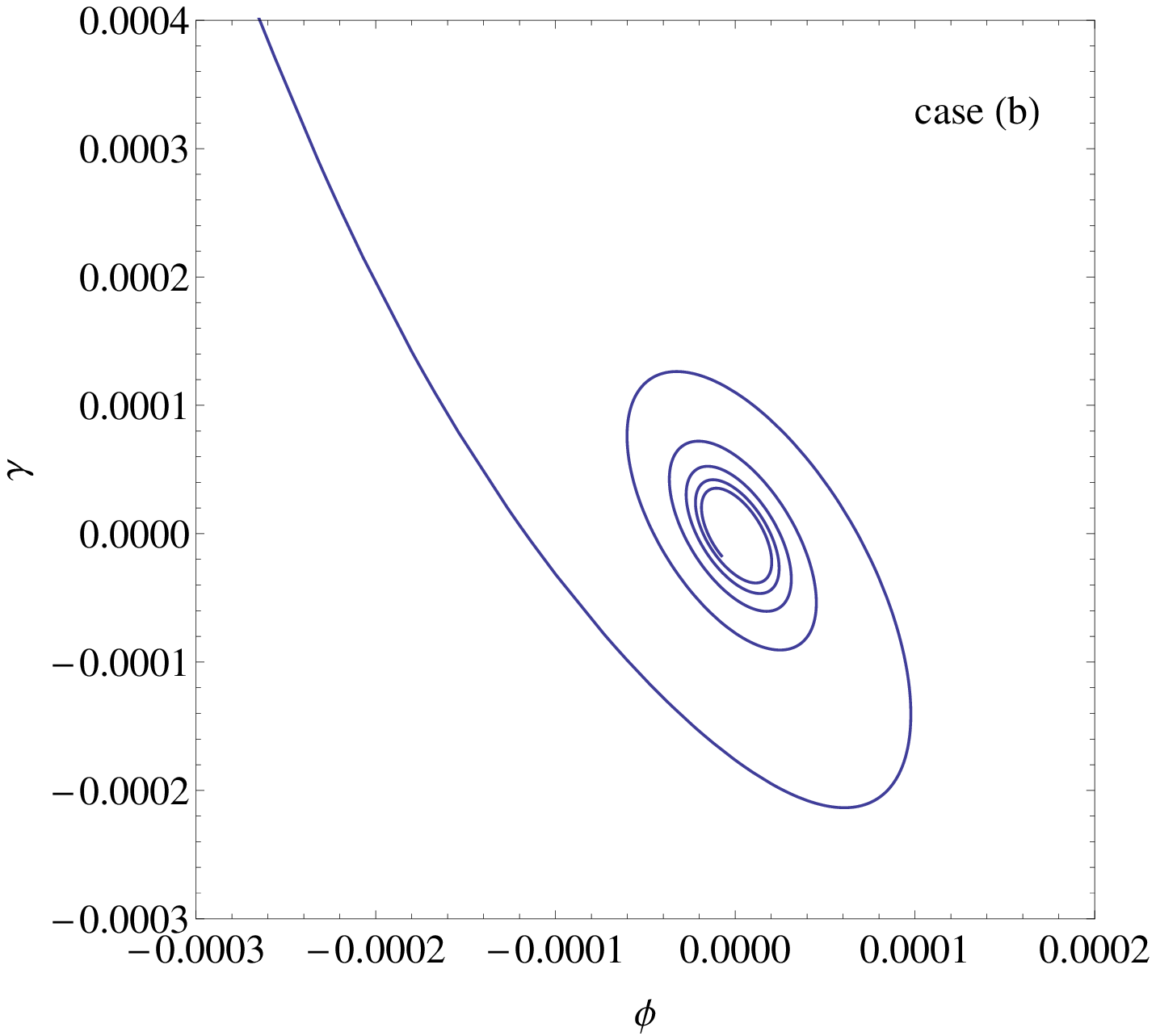}}\\[15pt]
	\scalebox{.7}{\includegraphics{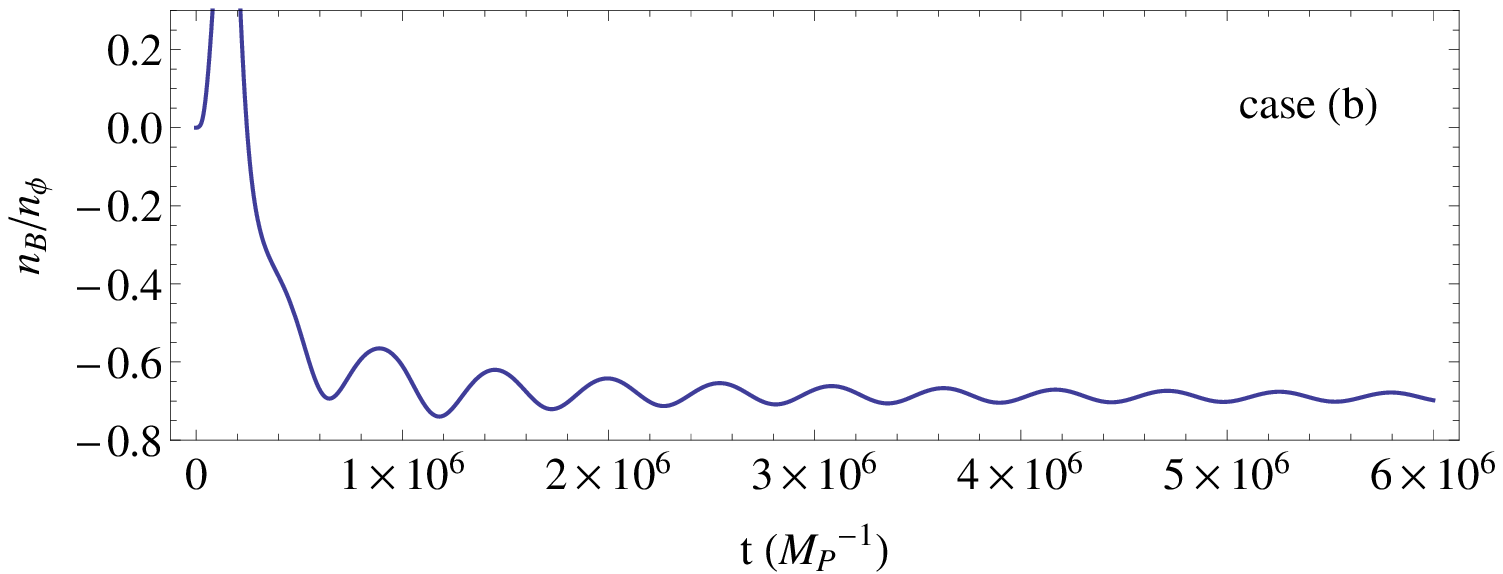}}
	\caption{Evolution of $\Phi$ in the complex plane for $H\leq m_{3/2}$, and baryon asymmetry, in case (b). The decay of $\Phi$ occurs at $t\simeq6.3\times10^6 M_P^{-1}$. }
\label{ADs2}
\end{figure}

When $H\sim m_{3/2}$, the effective mass of the AD field becomes positive, and $\Phi$ oscillates about the origin. The oscillations of $\phi$ and $\gamma$ are clearly out of phase in both cases, so $\Phi$ is kicked into a spiraling motion in the complex plane, as demonstrated by Figures  \ref{ADs1} and \ref{ADs2} corresponding to cases a) and b) respectively. It is in this spiral motion that the baryon number of the flat direction is stored. The lower panels of Figures \ref{ADs1} and \ref{ADs2} demonstrate how the ratio $n_B/n_{\phi}$ asymptotes to a constant value, dependent on the amount of $CP$ violation, but independent of the masses of the inflaton and the AD field. Further numerical calculations show this ratio to be relatively independent of the size of the parameter $\lambda/M$.

\section{Summary and conclusion}

In this paper we have presented a possible mechanism to generate a large vacuum expectation value for a flat direction during inflation, necessary to realize Affleck-Dine baryogenesis. In this scenario, the energy density of the universe is always dominated by the inflaton and its decay products, allowing for a simple estimate of the final baryon asymmetry produced after the decay of the AD field.

In the chaotic inflation model studied here, the stabilizer field $S$ is necessary to lift the flatness of the potential for the inflaton $I$. The inflationary potential can then be chosen by fixing the inflaton dependence of the superpotential $W=Sf(I)$. To eliminate the inflationary perturbations of $S$, a non-minimal K\"ahler term of the form $-\xi(S\bar{S})^2$ is introduced. This in turn drives $S$ to zero, up to perturbations proportional to $\xi^{-1}$, which reduces the investigation to the dynamics of the inflaton and the AD field.  In the present study, the freedom of choice for the inflaton potential was not fully exploited, and we have specialized to the choice $f(I)=m_{\eta}I$ for simplicity. However, it must be noted that for any choice of $f$ that supports slow roll inflation, the addition of an Affleck-Dine and supersymmetry breaking sectors does not perturb the classical evolution of the inflaton (see Appendix).

Spontaneous supersymmetry breaking was realized via a Polonyi-like sector with the usual Polonyi superpotential, and a non-minimal K\"ahler term with a stabilizing term $-(Z\bar{Z})^2/\Lambda^2$. The finite energy density during inflation induces a positive effective mass of the order of the Hubble parameter, which displaces the Polonyi field to an instantaneous minimum of order $(m_{3/2}/H)^2M_P$. After inflation, $Z$ evolves to the true supersymmetry breaking minimum. For small $\Lambda$, the non-minimal K\"ahler term damps the oscillations about this minimum, resolving the cosmological moduli and gravitino problems. Since the VEV of $Z$ at any time is always much smaller than the Planck scale, the evolution of the Affleck-Dine sector is insensitive to the dynamics of the supersymmetry breaking sector. The gravitino mass $m_{3/2}$ has an approximately constant value during the complete evolution of $Z$.

We have shown that the VEV necessary to drive the Affleck-Dine mechanism can be obtained via a non-minimal coupling between the scalar field parametrizing the flat direction, and the same stabilizer field $S$. The main virtue of this model lies in that it is generic; the instantaneous minimum is independent of the explicit form of the inflaton superpotential, since it only depends on $\eta$ through the Hubble parameter $H$. Therefore, the discussion of the evolution of $\Phi$ can be easily adapted for a wide range of models of chaotic inflation. Since most of the fields involved in the model are expected to vanish or remain very small, the relatively complex system of evolution equations can be integrated numerically for a relatively wide range of parameters. Despite the fact that this range does not include the more realistic values, the qualitative behavior of the solutions agrees with the approximate expressions derived in the paper.

\section*{Appendix}
In this appendix, we compute the perturbations due to the 
AD and Polonyi fields on the dynamics of the inflationary sector. Since these perturbations are expected to be small, they can be calculated expanding the complete scalar potential to quadratic order, and solving for the displacement that minimizes the potential. Schematically, if $\varphi_0$ is the unperturbed minimum for the field $\varphi$, then
\beq
V(\delta\varphi)\simeq V(\varphi_0)+V'(\varphi_0)\delta\varphi+\frac{1}{2}V''(\varphi_0)\delta\varphi^2,
\eeq
and the perturbation $\delta\varphi$ is such that $V'(\delta\varphi)=0$.

The reduced Planck mass, $M_P$, is again set to unity. For generality let us consider the superpotential (\ref{w_inf}), along with the slow roll parameters
\begin{equation}
\epsilon \equiv \frac{1}{2}\left(\frac{\partial_{\eta}V}{V}\right)^2=\left(\frac{\partial_{I}f}{f}\right)^2\ , \quad \tilde{\eta} \equiv \frac{\partial^2_{\eta}V}{V}=\frac{\partial^{2}_{I}f}{f}+\left(\frac{\partial_{I}f}{f}\right)^2\ ,
\end{equation}
and
\begin{equation}
\Delta \equiv \frac{m_{3/2}}{f} = \frac{m_{3/2}}{\sqrt{3}H_{I}}\ .
\end{equation}
We will only keep terms larger than $\mathcal{O}(|\Phi_0|^3,\Delta |\Phi_0|,\Delta^2)$, since during inflation it is expected that $|\Phi_0|^2<1$, $\Delta\ll 1$. For example, for $f(I)=m_{\eta}I$, 
\beq
\begin{aligned}
\epsilon&=\tilde{\eta}=5\times10^{-3}\left(\frac{20}{\eta}\right)^2, \\
 |\Phi_0|^2&=2\times10^{-5}(\zeta-1)^{1/2}\frac{\tilde{M}}{|\lambda|}\left(\frac{m_{\eta}}{10^{-5}}\right)\left(\frac{\eta}{20}\right), \\
\Delta &= 7\times10^{-12} \left(\frac{m_{3/2}}{10^{-15}}\right)\left(\frac{10^{-5}}{m_{\eta}}\right)\left(\frac{20}{\eta}\right).
\end{aligned}
\eeq
All first derivatives of the complete scalar potential (\ref{sugra_potential}) along the imaginary directions vanish. For the real parts,
\begin{equation}\label{1_ders}
\left(
\begin{matrix}
\partial_{\eta}\\
\partial_s\\
\partial_{\phi}\\
\partial_z
\end{matrix}
\right) V(\eta,0,\phi,0) \simeq \left(
\begin{matrix}
\sqrt{2\epsilon}[1-(\zeta-1)\Phi_0^2]\\
-2\sqrt{2}\Delta-\sqrt{\frac{\zeta-1}{6}}(2\zeta-1)\Phi_0^2\\
0\\
0
\end{matrix}
\right)f^2.
\end{equation}

The second derivatives are block diagonal; there is no mixing between real and imaginary parts. The second derivative matrices are shown in the next page. From these results it is immediate that, to this order, the imaginary parts of $I$ and $S$ will be unperturbed, and therefore should remain vanishingly small. Equations (\ref{1_ders}) and (\ref{2_ders}) show that the perturbation to the inflaton $\eta$ is negligible when $\epsilon\ll 1$, $\delta\eta\propto\sqrt{\epsilon}$. Thus, $\eta$ will be unperturbed during the inflationary phase, and it is expected that the chaotic inflation scenario will be realized. For the stabilizer $S$, the perturbation can be calculated as
\begin{equation}
\delta s = \frac{2\sqrt{2}\Delta}{\epsilon+4\xi}  +  \sqrt{\frac{\zeta-1}{6}}\left(\frac{2\zeta-1}{\epsilon+4\xi}\right)\Phi_0^2 + \mathcal{O}(\Phi_0^4).
\end{equation}
The first term will be negligible during inflation, $\Delta\ll 1$. However, the second term might be relatively large, depending on the value of $\zeta$ and the VEV acquired by the AD field. Since $s$ will not remain at zero, it will perform coherent oscillations at the end of inflation. For a sufficiently large $\xi$ this perturbation can be kept small. It must be emphasized that even if $S$ is perturbed, the inflaton $\eta$ will remain relatively unaffected.


\begin{landscape}

\begin{equation}\label{2_ders}
\begin{aligned}
\partial_{\eta,s,\phi,z}^2 V &= 
\left(
\begin{matrix}
\tilde{\eta} & -\Delta\sqrt{\epsilon} & 0 & 0 \\
-\Delta\sqrt{\epsilon} & \epsilon +4\xi & 0 & -\sqrt{3}\Delta \\
0 & 0 & 4(\zeta-1)+\sqrt{3(\zeta-1)}\delta & 0\\
0 & -\sqrt{3}\Delta & 0 &  1
\end{matrix}
\right) f^2\\
&\,\,\,\,+ 
\left(
\begin{matrix}
\tilde{\eta}(1-\zeta) & -\frac{1}{4}\sqrt{\epsilon\cdot\frac{\zeta-1}{3}}(2\zeta-1) & -2\sqrt{\epsilon}(\zeta-1)/\Phi_0 & 0 \\
-\frac{1}{4}\sqrt{\epsilon\cdot\frac{\zeta-1}{3}}(2\zeta-1) & \frac{2}{3}(\zeta^2-2\zeta+1)+\epsilon-8\xi(\zeta-\frac{1}{2}) & -2\sqrt{\frac{\zeta-1}{3}}(2\zeta-1)/\Phi_0 & 0 \\
-2\sqrt{\epsilon}(\zeta-1)/\Phi_0 & -2\sqrt{\frac{\zeta-1}{3}}(2\zeta-1)/\Phi_0   & 6\zeta^2+\frac{25}{4}\zeta-\frac{37}{4} & 0\\
0 & 0 & 0 & -\frac{2}{3}(\zeta-1)
\end{matrix}
\right) \Phi_0^2f^2\ .
\end{aligned}
\end{equation}\\

\begin{equation}
\begin{aligned}
\partial_{\beta,\alpha,\gamma,\chi}^2 V &= 
\left(
\begin{matrix}
2+2\epsilon - \tilde{\eta} & -(2+a)\Delta\sqrt{\epsilon} & 0 & 0\\
-(2+a)\Delta\sqrt{\epsilon} & \epsilon +4\xi & 0& -\sqrt{3}\Delta \\
0 & 0 & -\sqrt{3(\zeta-1)}\Delta & 0\\
0 & -\sqrt{3}\Delta & 0 & 1\\ 
\end{matrix}
\right)f^2\\
&\,\,\,\,+\left(
\begin{matrix}
(\tilde{\eta}-2\epsilon-\frac{4}{3})(\zeta-1) & -\sqrt{\epsilon\cdot\frac{\zeta-1}{3}}(\zeta+1) & 0 & 0\\
-\sqrt{\epsilon\cdot\frac{\zeta-1}{3}}(\zeta+1) & \frac{2}{3}\zeta(\zeta+4)+\epsilon-8\xi(\zeta-\frac{1}{2}) &2 \sqrt{\frac{\zeta-1}{3}}(\zeta+1)/\Phi_0 & 0 \\
0 & 2\sqrt{\frac{\zeta-1}{3}}(\zeta+1)/\Phi_0 & 2\zeta^2-\frac{1}{4}\zeta-\frac{3}{4}  & 0\\[5pt]
0 & 0 & 0 & -\frac{2}{3}(\zeta-1)
\end{matrix}
\right) \Phi_0^2f^2\ .
\end{aligned}
\end{equation}\\

\begin{center}
Second derivatives for the inflation+Polonyi+AD model
\end{center}

\end{landscape}

\section*{Acknowledgments}
We would like to thank A. Linde and M. Peloso for helpful discussions.
This work was supported in part
by DOE grant DE--FG02--94ER--40823 at the University of Minnesota.


\end{document}